\documentclass[aps,prd,groupaddress,floatfix,tighten,nofootinbib,showpacs,amsfonts,superscriptaddress]{revtex4-2}
\usepackage{url}
\usepackage{xcolor}
\usepackage{hyperref}
\usepackage[latin1]{inputenc}

\colorlet{shadecolor}{gray!15}
\definecolor{greenLinks}{rgb}{0, 0.6, 0}
\definecolor{blueLinks}{rgb}{0, 0, 0.6}
\definecolor{redLinks}{rgb}{0.6, 0, 0}
\definecolor{tempText}{rgb}{0.55, 0.10,0.67}
\definecolor{eprintLinks}{rgb}{0.4, 0.4, 0.4}
\definecolor{journalLinks}{rgb}{0.6, 0, 0}
\usepackage{amssymb}
\usepackage{amsmath,color}
\usepackage{epsfig}
\usepackage{graphicx,epsf,epsfig}
\usepackage{footnote}
\usepackage{multirow}
\usepackage{float}
\usepackage{enumitem}
\usepackage{color}
\def\slc#1{\setbox0=\hbox{$#1$}                  % set a box for #1
    \dimen0=\wd0                                 % and get its size
    \setbox1=\hbox{/} \dimen1=\wd1               % get size of /
    \ifdim\dimen0>\dimen1                        % #1 is bigger
       \rlap{\hbox to \dimen0{\hfil/\hfil}}      % so center / in box
       #1                                        % and print #1
    \else                                        % / is bigger
       \rlap{\hbox to \dimen1{\hfil$#1$\hfil}}   % so center #1
       /                                         % and print /
    \fi}

\def\be{\begin{equation}}
\def\ee{\end{equation}}
\def\gs{\mathrel{
   \rlap{\raise 0.511ex \hbox{$>$}}{\lower 0.511ex \hbox{$\sim$}}}}
\def\ls{\mathrel{
   \rlap{\raise 0.511ex \hbox{$<$}}{\lower 0.511ex \hbox{$\sim$}}}}

\newcommand{\ba}{\begin{array}{c}}
\newcommand{\baz}{\begin{array}{cc}}
\newcommand{\barrr}{\begin{array}{rrr}}
\newcommand{\bad}{\begin{array}{ccc}}
\newcommand{\bav}{\begin{array}{cccc}}
\newcommand{\baf}{\begin{array}{ccccc}}
\newcommand{\bea}{\begin{equation} \begin{array}{c}}
\newcommand{\eea}{\end{array} \end{equation}}
\newcommand{\ea}{\end{array}}

\usepackage[normalem]{ulem}

\def\21{$\mathrm{SU(2)_L \otimes U(1)_Y}$ }

\newcommand {\ignore}[1]{}

\newcommand{\vt}{\vert}
\newcommand{\nn}{\nonumber}

\allowdisplaybreaks \allowdisplaybreaks[2]
\newcommand{\AddrESFM}{Escuela Superior de Física y Matemáticas, Instituto Politécnico Nacional, Av. Instituto Politécnico Nacional s/n Edificio 9 Unidad Profesional ``Adolfo López Mateos'' Col. San Pedro Zacatenco, Alcald\'ia Gustavo A. Madero,  C.P. 07738, Ciudad de M\'exico; M\'exico.}

\newcommand{\AddrCECYT}{Centro de Estudios Cient\'ificos y Tecnol\'ogicos No 16, Instituto Polit\'ecnico Nacional, Kilómetro 1.500, Actopan - Pachuca, Distrito de Educaci\'on, Salud, Ciencia, Tecnolog\'ia e Innovaci\'on, San Agust\'in Tlaxiaca, Hidalgo; M\'exico. 
}

\begin{document}
%-----------------------------------------------------------------------------
\title{Soft breaking of the $\mu\leftrightarrow \tau$ symmetry by  $\mathbf{S}_{4}\otimes \mathbf{Z}_{2}$ } 
%-----------------------------------------------------------------------------
%
\author{J. D. Garc\'ia-Aguilar}
\email{jdgarcia@ipn.mx}
\affiliation{\AddrCECYT}

\author{Asahel Enrique Pozas Ramírez}
\email{apozasr1700@alumno.ipn.mx}
\affiliation{\AddrESFM}

\author{Marlon Michael Suárez Castañeda}
\email{msuarezc1700@alumno.ipn.mx}
\affiliation{\AddrESFM}

\author{Juan Carlos G\'omez-Izquierdo}
\email{cizquierdo@ipn.mx\\
}
\affiliation{\AddrCECYT}
%

%

%
%-----------------------------------------------------------------------------
%\pacs{14.60.Pq, 11.30.Er}

\date{\bf \today} 

\begin{abstract}\vspace{2cm}
The $\mu \leftrightarrow \tau$ symmetry has been ruled out by its predictions on the reactor and atmospheric angles, nevertheless, a breaking of this symmetry might provide correct values. For that reason, we build a non-renormalizable lepton model where the mixings arise from the spontaneous breaking of the $\mathbf{S}_{4}\otimes \mathbf{Z}_{2}$ discrete group, subsequently the $\mu \leftrightarrow \tau$ symmetry is broken in the effective neutrino mass matrix, that comes from the type II see-saw mechanism. As main result, the reactor and atmospheric angles are corrected and their values are in good agreement with the experimental data for the inverted hierarchy. Furthermore, we point out a link between the atmospheric angle and reactor one. In the quark sector, under certain assumptions, the generalized Fritzsch textures shape to the quark mass matrices so that the CKM matrix values are guaranteed.

%we find the values for the reactor angle and the atmospheric angle consistent with the experimental data taking into account the inverted neutrino hierarchy. Furthermore,  we observe a link between the atmospheric angle and reactor one.

%as a result the reactor and atmospheric angles are consistent with the experimental data for the inverted hierarchy. 
%As an outstanding result, the deviation of the atmospheric angle is linked with the reactor one. In the quark sector, under certain assumptions, the generalized Fritzsch textures shape to the quark mass matrices so that the CKM matrix values are guaranteed.}

%As an interesting result, the deviation of the atmospheric angle is related with the reactor one, this is,    $\sin{\theta_{23}}\approx  \frac{1}{\sqrt{2}} \vert 1\pm\frac{\sin{\theta_{13}}}{\sqrt{2}}\vert$. 

\end{abstract}

\begin{flushright}
CECyT 16-21-I
\end{flushright}

%-----------------------------------------------------------------------------
\maketitle
%-----------------------------------------------------------------------------
%

\section{Introduction}
The confirmed $(g-2)_{\mu}$ anomaly by the Fermilab~\cite{Muong-2:2021ojo} and the $W$ boson mass new measurement~\cite{CDF:2022hxs}, which are not consistent with the theoretical prediction, have shown again that the Standard Model (SM) is incomplete. Along with this, the flavor puzzle remains to be solved so that there is a need to enlarge the SM. 

The neutrino oscillation opened the window to search physics beyond the SM, as it is well known, these
established that neutrinos have mass so they mix. Although there are many mechanism~\cite{Romao:2007ny} to get tiny neutrino masses, so far there is no a convincing theory that explains the origin of such mass and the peculiar pattern  which is completely different from the quark sector. In the last years, several experiments have measured the neutrino mixing angles with great accuracy, also the masses seem to obey two orderings (normal and inverted hierarchy) due to the lacking of information on the absolute neutrino mass. Certainly, the normal ordering is preferred by the available data~\cite{deSalas:2020pgw, Esteban:2020cvm} but the inverted hierarchy is not completely discarded~\cite{Gariazzo:2022ahe}. Another important point is that, 
conforming to the experimental data, the PMNS mixing matrix exhibits large values in its entries, in addition, the second and third rows satisfy the relation $\vert \mathbf{U}_{\mu i}\vert=\vert \mathbf{U}_{\tau i}\vert$ ($i=1,2,3$) in good approximation for the normal and inverted hierarchy. The aforementioned facts
might be understood by means of a symmetry in the effective neutrino mass matrix, then the concept of flavor symmetry turn out being crucial to explain the mixings, and a variety of discrete symmetries~\cite{Ishimori:2010au, Grimus:2011fk, Ishimori:2012zz, King:2013eh, King:2015aea, Grimus:2011fk, Feruglio:2019ybq, Xing:2020ijf, Chauhan:2022gkz} have been applied to the lepton sector. In particular, the neutrino data seem to obey an approximated  $\mu \leftrightarrow \tau$ symmetry (for a complete review see~\cite{Xing:2015fdg}), that consists in the exchange label $\mu\leftrightarrow \tau$ in the effective neutrino mass matrix when the charged lepton mass one is diagonal. Speaking of exact $\mu \leftrightarrow \tau$ symmetry, which is is outdated currently due to its predictions, would imply to obtain
$0^{\circ}$ and $45^{\circ}$ for the reactor and atmospheric angles, respectively. Besides that, the solar angle and the Dirac CP-violating phase keep as a free and unknown parameters. Despite this, from the model building point of view,  the well studied $\mu \leftrightarrow \tau$ symmetry has been a guide to construct lepton models~\cite{Grimus:2003kq, Grimus:2004rj, Adulpravitchai:2008yp,Ishimori:2008gp, Hagedorn:2010mq, Gomez-Izquierdo:2017rxi, Garces:2018nar, Gomez-Izquierdo:2018jrx, Morisi:2010rk} and there is a possibility that a soft breaking ~\cite{Morisi:2011pm,Gupta:2011ct, Gupta:2013it,Zhao:2016orh, Gomez-Izquierdo:2017rxi, Garces:2018nar, Gomez-Izquierdo:2018jrx, Becerra-Garcia:2022gys} of this symmetry can accommodate the experimental results so that there is still strong motivation to study on the $\mu \leftrightarrow \tau$ symmetry. Apart from this, 
elaborated flavored models have been proposed to face the lepton mixings and related issues as leptogenesis, dark matter, and so forth~\cite{Xing:2020ald, Calibbi:2021qto, Chauhan:2022gkz}.

On the other hand, in the quark sector, according to the available data~\cite{Workman:2022ynf} the CKM matrix is close to the identity one,  this pattern might be explained by the notable hierarchy among the quark masses. In addition, this feature is exhibited by some matrices like the nearest neighbor interactions (NNI)~\cite{Branco:1988iq, Branco:1994jx,Harayama:1996am,Harayama:1996jr} and the generalized Fritzsch~\cite{Fritzsch:1999ee, Barranco:2010we, Fritzsch:2015gxa} mass textures which can be obtained by means the flavor symmetries~\cite{Ishimori:2010au, Grimus:2011fk, Ishimori:2012zz, King:2013eh}. The contrasting behavior between the PMNS and CKM mixing matrices is undoubtedly a puzzling problem, so far one of the main task for model builders is to match simultaneously the fermion mixings by the same flavor symmetry in the suitable framework.

In order to address the masses and mixing problem, a phenomenological scalar extension of the SM is realized where the type II see-saw mechanism is responsible to obtain small neutrino masses and special emphasis is put on the lepton sector under a soft breaking of the $\mu \leftrightarrow \tau$ symmetry scheme. To do so, we use the $\mathbf{S}_{4}$\cite{Patel:2010hr, Dong:2010zu,Ishimori:2010fs,Mohapatra:2012tb,BhupalDev:2012nm,VanVien:2015xha,Vien:2019zhs,Vien:2020aya,Vien:2022pwf} non-abelian discrete group to handle the Yukawa couplings, at the same time, this symmetry allows to treat the quark, lepton and scalar sector in different manner. Additionally, we include  a $\mathbf{Z}_{2}$  symmetry, to have a non-renormalizable Yukawa mass term for neutrinos. On the other hand,  
the inclusion of three Higgs doublets are required to obtain the quark and charged lepton masses and mixings, this latter comes out being diagonal as result of the matter assignation under the flavor symmetry. Then, an enriched scalar (flavons) sector is included to provide desirable mass textures. In consequence, the mixings arises from the spontaneous breaking of the $\mathbf{S}_{4}\otimes \mathbf{Z}_{2}$ discrete group and the $\mu \leftrightarrow \tau$ symmetry is broken in the effective neutrino mass matrix. Eventually, the reactor and atmospheric angles come out being different of $0^{\circ}$ and $45^{\circ}$ respectively.
CP parities phases in the neutrino masses play an important role to get sizable values for $\theta_{13}$ and the deviation of $\theta_{23}$ from maximality which turn out being consistent with neutrino data for the inverted hierarchy. In the quark sector, under certain assumptions, the generalized Fritzsch textures shape to the quark mass matrices so that the CKM matrix values are guaranteed.

It is worthy mentioned that a similar study was carried out~\cite{Morisi:2010rk}, nonetheless there are clear differences namely. The first one is scalar matter and the flavor symmetry, the second one is related with the mechanism to generate small neutrino masses and the corresponding predictions: in the aforementioned paper, they got exact $\mu \leftrightarrow \tau$ symmetry. Lastly, the NNI textures, in the quark mass matrices, appeared in a natural way so that they obtained correct values for the mixings. Although our model has some limitations like the flavon  alignments and one benchmark (in the quark sector), the main purpose of this work was to show that a simple soft breaking of the $\mu \leftrightarrow \tau$ symmetry is enough to correct the lepton mixing angles.

The layout of the paper is as follows. In section II, we describe the general framework to explore the $\mathbf{S}_{4}$ discrete symmetry, the full assignation for the matter content is shown and the mass matrices and the corresponding mixing matrix are obtained. In addition, a brief analytical study is carried out to fix some free parameters in the model. Main results are presented in scattered plots where the set of free parameters values, that fit the mixing angles, are shown. All of this is included in section III. We give some conclusions in section IV.

\section{Flavored Model}

\subsection{General framework}

Although, there are fascinating theoretical frameworks that can be good candidates to replace the SM, conforming to our interest, a scalar extension of the SM will be considered. Thus, apart from the SM matter content a Higgs triplet ($\Delta$) is required to generate tiny neutrino masses by means the type II see-saw mechanism. Furthermore, extra Higgs doublets and flavon  gauge singlets ($\phi$, $\varphi$ and $\xi$) will be added to provide the CKM and PMNS matrices, respectively. In Table~\ref{SMC1}, we can see the rest of the matter fields.

% the matter content will be considered to study masses and mixings, to generate tiny neutrino masses, in this simple extension of the SM, the type II see-saw mechanism is required. In consequence a Higgs triplet ($\Delta$) will be included in the matter content as one can see in Table~\ref{SMC1}.

\begin{table}[ht]
	\begin{center}
		\begin{tabular}{|c|c|c|c|c|c|c|c|c|c|c|}
			\hline\hline
			{\footnotesize Matter}	& {\footnotesize $Q_{L}=
				\begin{pmatrix}
					u \\
					d \\
				\end{pmatrix}_{L}$ } & {\footnotesize $d_{R}$} & {\footnotesize $u_{R}$} &  {\footnotesize  $L=
			\begin{pmatrix}
			\nu \\
			\ell \\
		\end{pmatrix}$ } & {\footnotesize $e_{R}$} & {\footnotesize $
	H=\begin{pmatrix}
	H^{+}\\ H^{0}
\end{pmatrix}$ } & {\footnotesize  $\Delta_{L}=\begin{pmatrix}
\frac{\delta^{+}}{2} & \delta^{++} \\
\delta^{0} & -\frac{\delta^{+}}{2} \\
\end{pmatrix}$}\\\hline
			{\footnotesize ${\bf SU(3)}_{c}$}	& {\footnotesize  $3$} & {\footnotesize  $3$} & {\footnotesize \bf $3$} & $1$ & $1$ & 1 & 1\\ \hline
			{\footnotesize ${\bf SU(2)}_{L}$}	& {\footnotesize \bf $2$} & {\footnotesize \bf $1$} & {\footnotesize \bf $1$} & $2$ & $1$ & $2$ & $3$\\ \hline
			{\footnotesize  ${\bf U(1)_{Y}}$}	& $\frac{1}{3}$ & -$\frac{2}{3}$ & $\frac{4}{3}$ & $-1$ & $-2$ & $1$ & $2$ \\\hline\hline
		\end{tabular}\caption{Matter content}\label{SMC1}
	\end{center}
\end{table}

%Another point to consider is the inclusion of extra Higgs doublets and flavon  gauge singlets ($\phi$, $\varphi$ and $\xi$) to obtain the CKM and PMNS matrices, respectively. This issue will be discussed in the next section. 
The relevant gauge invariant Lagrangian is given by
\begin{equation}
-\mathcal{L}=y^{d}\bar{Q}_{L}H d_{R}+y^{u}\bar{Q}_{L}\tilde{H} u_{R}+y^{e}\bar{L}H e_{R}+ \frac{1}{2}y^{\nu}\bar{L}(i\sigma_{2})\Delta \left(L\right)^{c}+V(H,\Delta, \phi,\varphi,\xi)+h.c.\label{eq1}
\end{equation}
with $\tilde{H}=i\sigma_{2}H^{\ast}$ and the scalar potential	
\begin{eqnarray}
V(H,\Delta, \phi,\varphi,\xi)&=&m^{2}_{H}H^{\dagger} H+\frac{1}{2}\lambda_{H}\left(H^{\dagger} H\right)^{2}+m^{2}_{\Delta} \textrm{Tr}(\Delta^{\dagger} \Delta)+\frac{1}{2}\lambda_{\Delta}\left(\textrm{Tr}(\Delta^{\dagger} \Delta)\right)^{2}+\lambda_{H\Delta}\left(H^{\dagger} H\right)\textrm{Tr}\left(\Delta^{\dagger} \Delta\right)\nonumber\\ &+&\lambda^{\prime}_{H\Delta}H^{T} \Delta^{\dagger} H
+m^{2}_{\phi} \vert \phi\vert^{2}+\frac{1}{2}\lambda_{\phi}\vert \phi\vert^{4}+\lambda_{H\phi}\left(H^{\dagger} H\right)\vert \phi\vert^{2}+\lambda_{\Delta\phi}\textrm{Tr}\left(\Delta^{\dagger} \Delta\right)\vert \phi\vert^{2}+m^{2}_{\varphi} \vert \varphi\vert^{2}+\frac{1}{2}\lambda_{\varphi}\vert \varphi\vert^{4}\nonumber\\&+&\lambda_{H\varphi}\left(H^{\dagger} H\right)\vert \varphi\vert^{2}+\lambda_{\Delta\varphi}\textrm{Tr}\left(\Delta^{\dagger} \Delta\right)\vert \varphi\vert^{2}+m^{2}_{\xi} \vert \xi\vert^{2}+\frac{1}{2}\lambda_{\xi}\vert \xi\vert^{4}+\lambda_{H\xi}\left(H^{\dagger} H\right)\vert \xi\vert^{2}+\lambda_{\Delta\xi}\textrm{Tr}\left(\Delta^{\dagger} \Delta\right)\vert \xi\vert^{2}\nonumber\\&+&\lambda_{\varphi\phi} \vert \varphi\vert^{2}\vert \phi\vert^{2}+\lambda_{\xi\phi} \vert \xi\vert^{2}\vert \phi\vert^{2}+\lambda_{\xi\varphi} \vert \xi\vert^{2}\vert \varphi\vert^{2}. \label{scpt}
\end{eqnarray}

In flavored models, the scalar potential turns out being important to get a viable model, in here, a detailed study on the scalar potential is not the purpose of this paper however we add a comment about it. The $\mathbf{S}_{4}$ discrete symmetry~\cite{Ishimori:2010au, Grimus:2011fk, Ishimori:2012zz, King:2013eh} was
selected to control the flavor mixings since it has singlet, doublet and triplet irreducible representations (see the appendix A for more details), this feature represents an advantage for us because the quark and Higgs sector will be assigned in doublets and singlets whereas the lepton sector in triplets. The main achievement to do that is to get desirable mass textures in both sectors.

Along with this, we wish to highlight the scalar potential, where the three Higgs doublets are only involved, has been study exhaustively ~\cite{Pakvasa:1977in, Beltran:2009zz, Das:2014fea}. In the aforementioned paper, three Higgs doublets were assigned under the $\mathbf{S}_{3}$ group as follows:
the first and second family were put in a $\mathbf{2}$ whereas the third one in $\mathbf{1}_{1}$. In these circumstances the scalar potential was minimized and the $\langle H_{2}\rangle=\langle H_{1}\rangle$ alignment is allowed by the flavor symmetry. Having commented that, we go back to our work where the $\mathbf{S}_{4}$ flavor symmetry drives the Yukawa couplings as well as the scalar potential. It is worthy mentioned that the non-abelian groups 
 $\mathbf{S}_{4}$ and $\mathbf{S}_{3}$ are completely different from each other ($\mathbf{S}_{3}$ is a subgroup of the $\mathbf{S}_{4}$), however, the $\mathbf{S}_{4}$ scalar potential with three Higgs doublets can be mimicked from the previous study~\cite{Pakvasa:1977in, Beltran:2009zz, Das:2014fea}. This asseveration is supported due to the $\mathbf{S}_{4}$ representation can be decomposed in the $\mathbf{S}_{3}$ ones~\cite{Ishimori:2010au}. To be more explicit, as  we can see in the appendix B, the irreducible representations $\mathbf{1_{1}}$, $\mathbf{1_{2}}$ and $\mathbf{2}$ of both groups coincide so that the tensor product respects the same rules among them as can be verified. Then, in this sense, similar results are expected for the Higgs alignments because we are using the same assignation for the three Higgs families under the $\mathbf{S}_{4}$, as one can see in Table~\ref{fma}. On the other hand, a complete analysis of the scalar potential is beyond the scope of this work so that the flavor alignments will be considered as a matter of fact.

Further to our previous comments, the full symmetry breaks down as follows: $\mathbf{SU(3)}_{C}\otimes \mathbf{SU(2)}_{L}\otimes \mathbf{U(1)}_{Y}\otimes \mathbf{S}_{4}\otimes \mathbf{Z}_{2}$ $\rightarrow$ $\mathbf{SU(3)}_{C}\otimes \mathbf{SU(2)}_{L}\otimes \mathbf{U(1)}_{Y}$ $\rightarrow$ $\mathbf{SU(3)}_{C}\otimes  \mathbf{U(1)}_{Q}$, 
 where the $\Lambda$ scale of the spontaneous breaking of the $\mathbf{S}_{4}\otimes \mathbf{Z}_{2}$ group is larger than the $v=246~GeV$ electroweak one.

%this is, under the $\mathbf{S}_{4}$ flavor symmetry those Higgs doublets are assigned as follows: the first and second family  are put in a $\mathbf{2}$ whereas the third one in $\mathbf{1}_{1}$ (see Table \ref{fma}). With this assignation, the scalar potential can be written explicitly
%we will not write explicitly the flavored invariant potential because this was obtained under the $\mathbf{S}_{3}$ flavor symmetry \cite{Das:2014fea}, which are a subgroup of the $\mathbf{S}_{4}$\footnote{To a description of $\mathbf{S}_{4}$, see appendix B.}. 
%In this sense, we expect similar results so some  alignments in the vacuum expectation values (vev's) for the Higgs fields and the flavons will be done.

%Remarkably, the $\mathbf{S}_{4}$ non-abelian group has been utilized to obtain the tribimaximal matrix~\cite{Harrison2002167, Xing200285,Altarelli:2012ss} and the nearest neighbor interaction (NNI) textures in the lepton and quark sector~\cite{Morisi:2010rk}, respectively. 
  
\subsection{The model}

Having commented briefly the theoretical framework, we put now attention to the matter field assignation under the $\mathbf{S}_{4}$ flavor symmetry. Hence, those are assigned as follows: the first and second family of quark and Higgs are put in $\mathbf{2}$ doublet; the third family is assigned to the $\mathbf{1}_{1}$ singlet. This choice has been exploited in many $\mathbf{S}_{3}$ models with three Higgs doublets (see for instance~\cite{Canales:2013cga}) and interesting mass textures can be obtained, for this reason, the same assignation is used in our work. On the other hand, the lepton sector is treated in different way since the three families of $L$ ($e_{R}$) left-handed (right-handed) doublets (singlets) are put in the $\mathbf{3}_{1}$ triplet irreducible representations. This allows to obtain a diagonal charged lepton mass matrix so that the mixings will arise from the neutrino sector where an enriched scalar one is needed as can be seen in Table~\ref{fma}. Let us add a comment on the role $\mathbf{Z}_{2}$ symmetry, the main purpose is to prohibit the renormalizable neutrino mass term, $\bar{L}(i\sigma_{2})\Delta L^{C}$.

\begin{table}[ht]
	\begin{center}
		\begin{tabular}{|c|c|c|c|c|c|c|c|c|c|c|c|c|c|c|}
\hline\hline
{\footnotesize Matter} & {\footnotesize $Q_{I L}$} & {\footnotesize $Q_{3 L}$} & {\footnotesize $d_{I R}$} & {\footnotesize $d_{3 R}$} & {\footnotesize $u_{I R}$} & {\footnotesize $u_{3 R}$} & {\footnotesize $L_{i}$} & {\footnotesize $e_{i R}$} & {\footnotesize $H_{I}$}  & {\footnotesize $H_{3}$} & {\footnotesize $\Delta$} & {\footnotesize $\phi$} & {\footnotesize $\varphi_{I}$} & {\footnotesize $\xi_{i}$} \\\hline
{\footnotesize $\mathbf{S}_{4}$} & {\footnotesize \bf $2$} & {\footnotesize \bf $1_{1}$} & {\footnotesize \bf $2$} & {\footnotesize \bf $1_{1}$} & {\footnotesize \bf $2$} & {\footnotesize \bf $1_{1}$} & {\footnotesize \bf $3_{1}$} & {\footnotesize \bf $3_{1}$} & {\footnotesize \bf $2$} & {\footnotesize \bf $1_{1}$} & {\footnotesize \bf $1_{1}$} & {\footnotesize \bf $1_{1}$} & {\footnotesize \bf $2$} & {\footnotesize \bf $3_{1}$}\\\hline
{\footnotesize  $\mathbf{Z}_{2}$} & 1 & 1 & 1 & 1 & 1 & 1 & 1 & 1 & 1 & 1 & -1 & -1 & -1 & -1 \\\hline\hline
		\end{tabular}\caption{Assignment under $\mathbf{S}_{4}$ flavor group. Here, $I=1,2$ and $i=1,2,3$.}\label{fma}
	\end{center}
\end{table}

Consequently, the most relevant terms which are flavor and gauge invariant are written as\footnote{ We ought to comment there are terms as for example  $(\bar{L}_{I} L^{C}_{I})_{1_{1}}(\tilde{H}^{2}_{I})_{1_{1}}/ \Lambda^{2} $, $(\bar{L}_{I} L^{C}_{I})_{2}(\tilde{H}^{2}_{I})_{2}/ \Lambda^{2} $ and  $(\bar{L}_{I} L^{C}_{I})_{1_{1}}(\tilde{H}^{2}_{3})_{1_{1}}/ \Lambda^{2} $ which are invariant under all symmetries but these contributions are very subleading due to the hierarchy $\langle \Delta \rangle\ll v\ll  \Lambda$.} 
\begin{eqnarray}
	\mathcal{L}&=&y_{1}^{d}\left[ \bar{Q}_{1L}\left(
	H_{1}d_{2R}+H_{2}d_{1R}\right) +\bar{Q}_{2L}\left(
	H_{1}d_{1R}-H_{2}d_{2R}\right) \right] +y_{2}^{d}\left[ \bar{Q}%
	_{1L}H_{3}d_{1R}+\bar{Q}_{2L}H_{3}d_{2R}\right] +y_{3}^{d}\left[ \bar{Q}%
	_{1L}H_{1}+\bar{Q}_{2L}H_{2}\right] d_{3R}  \nn\\
	&&+y_{4}^{d}\bar{Q}_{3L}\left[ H_{1}d_{1R}+H_{2}d_{2R}\right] +y_{5}^{d}\bar{%
		Q}_{3L}H_{3}d_{3R}+y_{1}^{u}\left[ \bar{Q}_{1L}\left(
	\tilde{H}_{1}u_{2R}+\tilde{H}_{2}u_{1R}\right) +\bar{Q}_{2L}\left(
\tilde{H}_{1}u_{1R}-\tilde{H}_{2}u_{2R}\right) \right]\nn\\
&&+y_{2}^{u}\left[ \bar{Q}%
_{1L}\tilde{H}_{3}u_{1R}+\bar{Q}_{2L}\tilde{H}_{3}u_{2R}\right] +y_{3}^{u}\left[ \bar{Q}%
_{1L}\tilde{H}_{1}+\bar{Q}_{2L}\tilde{H}_{2}\right] u_{3R}+  y_{4}^{u}\bar{Q}_{3L}\left[ \tilde{H}_{1}u_{1R}+\tilde{H}_{2}u_{2R}\right] +y_{5}^{u}\bar{%
	Q}_{3L}\tilde{H}_{3}u_{3R}
\nn\\&&+y^{e}_{1}\left[\bar{L}_{1}H_{3}e_{1 R}+\bar{L}_{2}H_{3}e_{2 R}+\bar{L}_{3}H_{3}e_{3 R}\right]+y^{e}_{2}\left[\bar{L}_{1}H_{2}e_{1 R}-\frac{1}{2}\bar{L}_{2}\left(\sqrt{3}H_{1}+H_{2}\right)e_{2R}+\frac{1}{2}\bar{L}_{3}\left(\sqrt{3}H_{1}-H_{2}\right)e_{3 R}\right]\nn\\
	&&+y^{N}_{1}\left[\bar{L}_{1} (i\sigma_{2})\Delta \phi L^{C}_{1}+\bar{L}_{2}(i\sigma_{2})\Delta \phi L^{C}_{2}+\bar{L}_{3}(i\sigma_{2})\Delta \phi L^{C}_{3}\right]\frac{1}{\Lambda}\nn\\&&+y^{N}_{2}\left[\bar{L}_{1}(i\sigma_{2})\Delta \varphi_{2} L^{C}_{1}-\frac{1}{2}\bar{L}_{2}(i\sigma_{2})\Delta \left(\sqrt{3}\varphi_{1}+\varphi_{2}\right)L^{C}_{2}+\frac{1}{2}\bar{L}_{3}(i\sigma_{2})\Delta \left(\sqrt{3}\varphi_{1}-\varphi_{2}\right) L^{C}_{3}\right]\frac{1}{\Lambda}\nn\\&&+y^{N}_{3}\left[\bar{L}_{1}(i\sigma_{2})\Delta\left(\xi_{2}L^{C}_{3}+\xi_{3}L^{C}_{2}\right)+\bar{L}_{2}(i\sigma_{2})\Delta \left(\xi_{1}L^{C}_{3}+\xi_{3}L^{C}_{1}\right) +\bar{L}_{3}(i\sigma_{2})\Delta \left(\xi_{1}L^{C}_{2}+\xi_{2}L^{C}_{1}\right)\right]\frac{1}{\Lambda}+h.c.
\end{eqnarray}

Once the scalar fields get their vev's, the fermion masses are written as 
\begin{eqnarray}
 {\bf \mathcal{M}}_{q}&=& \begin{pmatrix}
	y^{q}_{2} \langle H_{3}\rangle+y^{q}_{1} \langle H_{2}\rangle	& y^{q}_{1} \langle H_{1}\rangle & 	y^{q}_{3} \langle H_{1}\rangle \\
	y^{q}_{1} \langle H_{1}\rangle & y^{q}_{2} \langle H_{3}\rangle-y^{q}_{1} \langle H_{2}\rangle & y^{q}_{3} \langle H_{2}\rangle \\
	y^{q}_{4} \langle H_{1}\rangle 	& y^{q}_{4} \langle H_{2}\rangle & y^{q}_{5} \langle H_{3}\rangle 
\end{pmatrix};\nn\\
\mathbf{\mathcal{M}}_{e}&=&\begin{pmatrix}
y^{e}_{1} \langle H_{3}\rangle+y^{e}_{2} \langle H_{2}\rangle	& 0 & 0 \\
0 & y^{e}_{1} \langle H_{3}\rangle -\frac{1}{2}y^{e}_{2}\left(\sqrt{3}\langle H_{1}\rangle+\langle H_{2}\rangle\right) & 0 \\
0	& 0 & y^{e}_{1} \langle H_{3}\rangle +\frac{1}{2}y^{e}_{2}\left(\sqrt{3}\langle H_{1}\rangle-\langle H_{2}\rangle\right)
\end{pmatrix};\nn\\
\mathbf{\mathcal{M}}_{\nu}&=&\begin{pmatrix}
y^{N}_{1} \langle \phi\rangle+y^{N}_{2} \langle \varphi_{2}\rangle	& y^{N}_{3}\langle \xi_{3}\rangle & y^{N}_{3}\langle \xi_{2}\rangle \\
y^{N}_{3}\langle \xi_{3}\rangle & y^{N}_{1} \langle \phi\rangle-\frac{1}{2}y^{N}_{2}\left(\sqrt{3}\langle \varphi_{1}\rangle+\langle\varphi_{2}\rangle \right) & y^{N}_{3}\langle \xi_{1}\rangle \\
y^{N}_{3}\langle \xi_{2}\rangle & y^{N}_{3}\langle \xi_{1}\rangle & y^{N}_{1} \langle \phi\rangle+\frac{1}{2}y^{N}_{2}\left(\sqrt{3}\langle \varphi_{1}\rangle-\langle\varphi_{2}\rangle \right)
\end{pmatrix}\frac{\langle \Delta \rangle}{\Lambda}.\label{fmas1}
\end{eqnarray}

Evidently, there are too many free parameters in the fermion mass matrices however some ones can be reduce notably by making an alignment in the vev's of the scalar fields. In particular, $\langle H_{1}\rangle=\langle H_{2}\rangle$ will be assumed in the Higgs sector as we already commented. Also, Higgs vev's have to satisfy the relation $\sqrt{\langle H_{1}\rangle^{2}+\langle H_{2}\rangle^{2}+\langle H_{3}\rangle^{2}}=v=246~GeV$. For the flavon sector, the alignment that provides a good phenomenology in the neutrino mass matrix is given by
\begin{eqnarray}
	\langle \xi \rangle=\left(v_{\xi_{1}},v_{\xi}, v_{\xi}\right), \qquad \langle \phi \rangle= v_{\phi}\left(1, 1, 1\right), \qquad \langle \varphi\rangle= v_{\varphi}\left(1, 0\right).\label{fvev}
\end{eqnarray}
As it is usual, each vev's  of the flavons are set to be proportional to $\lambda \Lambda $ where $\lambda$ ($0.225$) is the Wolfenstein parameter and the cutoff scale of the model.

\subsection{Fermion masses and mixings}

\subsubsection{Lepton sector}

As was already commented, we put special emphasis on the lepton sector. To start with, let us focus in the charged lepton sector which is diagonal and the physical masses can be obtained straightforwardly. Nonetheless, a particular alignment was assumed, this is, $\langle H_{1}\rangle=\langle H_{2}\rangle$~\cite{Pakvasa:1977in, Beltran:2009zz, Das:2014fea} and the principal motivation has to do with the quark sector where outstanding mass textures appear. 

As consequence of the mentioned choice in the Higgs sector, the $y^{e}_{2}$ Yukawa coupling has to be negative and an extra rotations in the fields are necessary to obtain $\hat{\mathbf{M}}_{e}=\textrm{Diag.}\left(m_{e}, m_{\mu}, m_{\tau}\right)=\mathbf{U}^{\dagger}_{e L}\mathbf{\mathcal{M}}_{e}\mathbf{U}_{e R}$ with  $\mathbf{U}_{e(L, R)}=\mathbf{S}_{23}\mathbf{u}_{e(L, R)}$, therefore $\hat{\mathbf{M}}_{e}=\mathbf{u}^{\dagger}_{e L} \mathbf{m}_{e}\mathbf{u}_{e R}$ 
with 
\begin{equation}
\mathbf{m}_{e}=
\begin{pmatrix}
	y^{e}_{1} \langle H_{3}\rangle+y^{e}_{2} \langle H_{2}\rangle	& 0 & 0 \\
	0 & y^{e}_{1} \langle H_{3}\rangle +\frac{1}{2}y^{e}_{2}\left(\sqrt{3}-1\right)\langle H_{2}\rangle & 0 \\
	0	& 0 & y^{e}_{1} \langle H_{3}\rangle -\frac{1}{2}y^{e}_{2}\left(\sqrt{3}+1\right)\langle H_{2}\rangle
\end{pmatrix},\qquad \mathbf{S}_{23}=\begin{pmatrix}
1 & 0 & 0 \\
0 & 0 & 1 \\
0 & 1 & 0
\end{pmatrix}. \label{clmasses}
\end{equation}

From Eqn. \ref{clmasses}, one can identify the charged lepton masses
\begin{equation}
m_{e}=\vert	y^{e}_{1} \langle H_{3}\rangle+y^{e}_{2} \langle H_{2}\rangle \vert, \quad m_{\mu}=\vert y^{e}_{1} \langle H_{3}\rangle +\frac{1}{2}y^{e}_{2}\left(\sqrt{3}-1\right)\langle H_{2}\rangle \vert, \quad m_{\tau}=\vert y^{e}_{1} \langle H_{3}\rangle -\frac{1}{2}y^{e}_{2}\left(\sqrt{3}+1\right)\langle H_{2}\rangle \vert.\label{clmass}
\end{equation} 

We stress that there are few parameters to adjust the three charged lepton masses and this can be a weak point. This can be solved by including extra flavons however we want to keep the model simple so that this will not be carried out.

In the neutrino sector, on the other hand, due to phenomenological implications in the mass matrix we assume the alignments given in Eqn. (\ref{fvev})
%\begin{eqnarray}
%\langle \xi \rangle=\left(v_{\xi_{1}},v_{\xi}, v_{\xi}\right), \qquad \langle \phi \rangle= v_{\phi}\left(1, 1, 1\right), \qquad \langle \varphi\rangle= v_{\varphi}\left(1, 0\right).
%\end{eqnarray}
Along with this, in the standard basis, $\mathbf{\mathcal{M}}_{\nu}$ is diagonalized by the $\mathbf{U}_{\nu}$ matrix such that
$\hat{\mathbf{M}}_{\nu}=\textrm{Diag.}\left(m_{1}, m_{2}, m_{3}\right)=\mathbf{U}^{\dagger}_{\nu}\mathbf{\mathcal{M}}_{\nu}\mathbf{U}^{\ast}_{\nu}$ with $\mathbf{U}_{\nu}=\mathbf{S}_{23}\mathbf{u}_{\nu}$, then
$\hat{\mathbf{M}}_{\nu}=\mathbf{u}^{\dagger}_{\nu}\mathbf{m}_{\nu}\mathbf{u}^{\ast}_{\nu}$ where $\mathbf{S}_{23}$ has been shown before and $\mathbf{m}_{\nu}$ is given by
\begin{equation}
	\mathbf{m}_{\nu}=\begin{pmatrix}
		y^{N}_{1} v_{\phi} & y^{N}_{3} v_{\xi} & y^{N}_{3} v_{\xi} \\
		y^{N}_{3} v_{\xi} & y^{N}_{1}v_{\phi} +\frac{\sqrt{3}}{2}y^{N}_{2} v_{\varphi} & y^{N}_{3}v_{\xi_{1}} \\
		y^{N}_{3} v_{\xi} & y^{N}_{3}v_{\xi_{1}} & y^{N}_{1}v_{\phi} -\frac{\sqrt{3}}{2}y^{N}_{2} v_{\varphi}
	\end{pmatrix}\frac{\langle \Delta \rangle}{\Lambda}=\begin{pmatrix}
		m_{ee} & m_{e\mu} & m_{e\mu} \\
		m_{e\mu} & m_{\mu \mu} & m_{\mu \tau} \\
		m_{e\mu} & m_{\mu \tau} & m_{\tau \tau}
	\end{pmatrix}.
\end{equation}

Due to the charged lepton mass matrix is diagonal, one can identify clearly the physical masses, see Eqn.(\ref{clmass}). Therefore, in the effective mass matrix, $\mathbf{m}_{\nu}$, the $\mu \leftrightarrow \tau$ symmetry is broken because of the difference $m_{\mu\mu}\neq m_{\tau\tau}$ as one can notice. As result of this, the reactor and atmospheric angles will be deviated from $0^{\circ}$ and $45^{\circ}$, respectively. As it is usual, in the context of $\mu \leftrightarrow \tau$, the solar angle is a free parameter which can be fixed to the current experimental values but this will get correction since that  $m_{\mu\mu}\neq m_{\tau\tau}$.

In order to diagonalize the neutrino mass matrix, a perturbative analysis will be done in such a way that the matrix can be written as
\begin{equation}
\mathbf{m}_{\nu}=\overbrace{\begin{pmatrix}
	m_{ee} & m_{e\mu} & m_{e\mu} \\
	m_{e\mu} & m_{\mu \mu} & m_{\mu \tau} \\
	m_{e\mu} & m_{\mu \tau} & m_{\mu \mu}
\end{pmatrix}}^{\mathbf{m}^{0}_{\nu}}+\overbrace{\begin{pmatrix}
0 & 0 & 0 \\
0 & 0 & 0 \\
0 & 0 & m_{\mu \mu} \epsilon
\end{pmatrix}}^{\mathbf{m}^{\epsilon}_{\nu}},
\end{equation}

where the former matrix possesses exact $\mu \leftrightarrow \tau$ symmetry and it is broken in the latter one where
the dimensionless parameter $\epsilon\equiv \left(m_{\tau \tau}-m_{\mu \mu}\right)/m_{\mu\mu}$ has been defined and this quantify the breaking. As we observe, this can be written as $\epsilon\sim y^{N}_{2}/(y^{N}_{1}-y^{N}_{2})$ (vev's of the flavons are proportional to $\lambda\Lambda$). So that, if $y^{N}_{2}$ was zero, the $\mu \leftrightarrow \tau$ symmetry would be exact, then we assume that $\epsilon$  is small such that
this parameter will be treated as a perturbation, thus, a pertubative study at first order in $\epsilon$ will be carried out. Therefore, we demand
that $\vert \epsilon \vert\leq 0.3$ as consequence quadratic ($\vert \epsilon \vert^{2}$) terms will be neglected.

As a result of having a diagonal charged lepton mass matrix, there is no contribution to the mixings, then the neutrino sector will provide it. To see this, we go back to the $\mathbf{m}_{\nu}$ mass matrix where $\mathbf{m}^{0}_{\nu}$ is diagonalized by the following mixing matrix \footnote{See appendix B for a brief overview on $\mu \leftrightarrow \tau$ symmetry.}

\begin{equation}
\mathbf{U}^{0}_{\nu}= \begin{pmatrix}
\cos{\theta} & \sin{\theta}  & 0 \\
-\frac{\sin{\theta}}{\sqrt{2}} & \frac{\cos{\theta}}{\sqrt{2}}  & -\frac{1}{\sqrt{2}} \\
-\frac{\sin{\theta}}{\sqrt{2}} & \frac{\cos{\theta}}{\sqrt{2}}  & \frac{1}{\sqrt{2}}
\end{pmatrix},
\end{equation}

Hereafter, the superscripted in $\mathbf{U}^{0}_{\nu}$ and the matrix elements $m^{0}_{\alpha \beta}$ ($\alpha, \beta=e, \mu,\tau$), denotes quantities when the $\mu \leftrightarrow \tau$ symmetry is exact. 

Going back to the expression $\hat{\mathbf{M}}_{\nu}=\mathbf{u}^{\dagger}_{\nu}\mathbf{m}_{\nu}\mathbf{u}^{\ast}_{\nu}$, then $\mathbf{u}_{\nu}\approx \mathbf{U}^{0}_{\nu} \mathbf{U}^{\epsilon}_{\nu}$ which implies
\begin{equation}
\hat{\mathbf{M}}_{\nu}=\mathbf{U}^{\epsilon \dagger}_{\nu}\left[ \mathbf{U}^{0\dagger}_{\nu} \mathbf{m}^{0}_{\nu}\mathbf{U}^{0\ast}_{\nu} +\mathbf{U}^{0\dagger}_{\nu} \mathbf{m}^{\epsilon}_{\nu}\mathbf{U}^{0\ast}_{\nu}\right]\mathbf{U}^{\epsilon \ast}_{\nu},\quad \textrm{with} \quad \mathbf{U}^{0\dagger}_{\nu} \mathbf{m}^{0}_{\nu}\mathbf{U}^{0\ast}_{\nu}=\textrm{Diag.}\left(m^{0}_{1}, m^{0}_{2}, m^{0}_{3}\right).
\end{equation}

In addition, we have
\begin{equation}
\mathbf{U}^{0\dagger}_{\nu} \mathbf{m}^{\epsilon}_{\nu}\mathbf{U}^{0\ast}_{\nu}=
\frac{\epsilon~m^{0}_{\mu\mu}}{2}\begin{pmatrix}
	\sin^{2}{\theta} &-\frac{\sin{2\theta}}{2} & -\sin{\theta}\\
-\frac{\sin{2\theta}}{2} & \cos^{2}{\theta} & \cos{\theta} \\
-\sin{\theta}	& \cos{\theta} & 1
\end{pmatrix},\qquad m^{0}_{\mu\mu}=\frac{1}{2}\left(m^{0}_{1}\sin^{2}{\theta}+m^{0}_{2}\cos^{2}{\theta}-m^{0}_{3}\right).
\end{equation}

As we already commented, the parameter $\epsilon$ is considered as a perturbation so that the mixing matrix $\mathbf{U}^{\epsilon}_{\nu}$ is obtained by using perturbation theory\footnote{In Appendix C, we detail the process to figure out $\mathbf{U}^{\epsilon}_{\nu}.$} at first order in $\vert \epsilon\vert$. Consequently, we obtain 

\begin{equation}
\mathbf{U}^{\epsilon}_{\nu}\approx \begin{pmatrix}
N_{1} & 	-\frac{m^{0}_{\mu\mu}}{m^{0}_{2}-m^{0}_{1}}\frac{\sin{2\theta}}{4}\epsilon N_{2} & \frac{m^{0}_{\mu\mu}}{m^{0}_{1}-m^{0}_{3}}\frac{\sin{\theta}}{2}\epsilon N_{3}\\
	\frac{m^{0}_{\mu\mu}}{m^{0}_{2}-m^{0}_{1}}\frac{\sin{2\theta}}{4}\epsilon N_{1} & N_{2}  & -\frac{m^{0}_{\mu\mu}}{m^{0}_{2}-m^{0}_{3}}\frac{\cos{\theta}}{2}\epsilon N_{3}  \\
	-\frac{m^{0}_{\mu\mu}}{m^{0}_{1}-m^{0}_{3}}\frac{\sin{\theta}}{2}\epsilon N_{1} & \frac{m^{0}_{\mu\mu}}{m^{0}_{2}-m^{0}_{3}}\frac{\cos{\theta}}{2}\epsilon N_{2}  & N_{3}
\end{pmatrix},
\end{equation}
where the normalization factors are written as
\begin{eqnarray}
N_{1}&=&\left[1+\frac{\vert \epsilon\vert^{2} \sin^{2}{\theta}}{4}\left(\bigg|\frac{m^{0}_{\mu\mu}}{m^{0}_{1}-m^{0}_{3}} \bigg|^{2}+\cos^{2}{\theta}\bigg| \frac{m^{0}_{\mu\mu}}{m^{0}_{2}-m^{0}_{1}} \bigg|^{2}\right)\right]^{-1/2},\nonumber\\
N_{2}&=&\left[1+\frac{\vert \epsilon\vert^{2} \cos^{2}{\theta}}{4}\left(\bigg|\frac{m^{0}_{\mu\mu}}{m^{0}_{2}-m^{0}_{3}} \bigg|^{2}+\sin^{2}{\theta}\bigg|\frac{m^{0}_{\mu\mu}}{m^{0}_{2}-m^{0}_{1}} \bigg|^{2}\right)\right]^{-1/2},\nonumber\\
N_{3}&=&\left[1+\frac{\vert \epsilon\vert^{2} }{4}\left(\sin^{2}{\theta}\bigg|\frac{m^{0}_{\mu\mu}}{m^{0}_{1}-m^{0}_{3}} \bigg|^{2}+\cos^{2}{\theta}\bigg|\frac{m^{0}_{\mu\mu}}{m^{0}_{2}-m^{0}_{3}} \bigg|^{2}\right)\right]^{-1/2}.
\end{eqnarray}

At last, the theoretical formulas for the mixing angles are obtained by comparing our PMNS mixing matrix, $\mathbf{U}\approx \mathbf{U}^{\dagger}_{e}\mathbf{U}_{\nu}=\mathbf{U}^{0}_{\nu} \mathbf{U}^{\epsilon}_{\nu}$, with the standard parametrization, then we finally get
 
\begin{eqnarray}
\sin{\theta_{13}}&=& \big|\mathbf{U}_{13}\big|= \frac{ N_{3}}{4}  \bigg| \frac{m^{0}_{\mu\mu}\left(m^{0}_{2}-m^{0}_{1}\right)\epsilon}{\left(m^{0}_{2}-m^{0}_{3}\right)\left(m^{0}_{1}-m^{0}_{3}\right)}  \bigg| \sin{2\theta};\nonumber\\
\sin{\theta_{12}}&=&\frac{ \big|\mathbf{U}_{12}\big| }{\sqrt{1-\sin^{2}{\theta_{13}}}}=N_{2}\sin{\theta}\frac{\bigg| 1-\frac{\epsilon}{2}\left(\frac{m^{0}_{\mu\mu}}{m^{0}_{2}-m^{0}_{1}}\right)\cos^{2}{\theta}  \bigg| }{\sqrt{1-\sin^{2}{\theta_{13}}}};\nonumber\\
\sin{\theta_{23}}&=&\frac{\big|\mathbf{U}_{23}\big|}{\sqrt{1-\sin^{2}{\theta_{13}}}}=\frac{N_{3}}{\sqrt{2}} 
\frac{\bigg| 1+\frac{\epsilon}{2} \frac{m^{0}_{\mu\mu}\left(m^{0}_{2}\sin^{2}{\theta}+m^{0}_{1}\cos^{2}{\theta}-m^{0}_{3}\right)}{\left(m^{0}_{2}-m^{0}_{3}\right)\left(m^{0}_{1}-m^{0}_{3}\right)} \bigg|}{\sqrt{1-\sin^{2}{\theta_{13}}}}.
\end{eqnarray}
As one can realize if $\epsilon$ goes to zero, one would obtain the well known predictions: $\theta_{13}=0$, $\theta_{12}=\theta$ and $\theta_{23}=\dfrac{\pi}{4}$.

In order to figure out the set of free parameter values, an analytical study on the theoretical formulas is carried out. It is important to note that the reactor angle depends strongly on the breaking parameter and the ratio among complex masses, $m^{0}_{i}=\vert m^{0}_{i}\vert e^{i\alpha_{i}}$.
In the former factor, the associated phase $\epsilon=\vert \epsilon \vert e^{i\alpha_{\epsilon}}$ is irrelevant however the difference  $m^{0}_{2}-m^{0}_{1}$ and $m^{0}_{1}-m^{0}_{3}$ are crucial to enhance the reactor angle value, then CP parities values turn out being relevant to accommodate the reactor angle. As result of this, we choose  the following CP parities values $m^{0}_{2}=-\vert m^{0}_{2}\vert$,  $m^{0}_{1}=\vert m^{0}_{1}\vert$ and $m^{0}_{3}=\vert m^{0}_{3}\vert$. To add to it, the solar and atmospheric angles are sensitive to the associated phase $\alpha_{\epsilon}$ and the CP parities values of the neutrino masses.

In the current analysis, the normal hierarchy is not favored as one can check straightforward, then we just focus in the inverted ordering. Due to the CP parities in the neutrino masses, we obtain
\begin{eqnarray}
	\sin{\theta_{13}}&=& \big|\mathbf{U}_{13}\big|= \frac{ N_{3}}{8}  \bigg| \frac{\left[\left(\vert m^{0}_{2}\vert+\vert m^{0}_{1}\vert\right)\cos^{2}{\theta}-\left(\vert m^{0}_{1}\vert-\vert m^{0}_{3}\vert\right)\right]\left(\vert m^{0}_{2}\vert+\vert m^{0}_{1}\vert\right)}{\left(\vert m^{0}_{2}\vert+\vert m^{0}_{3}\vert\right)\left(\vert m^{0}_{1}\vert-\vert m^{0}_{3}\vert\right)}  \bigg| \vert \epsilon\vert \sin{2\theta};\nonumber\\
	\sin{\theta_{12}}&=&\frac{ \big|\mathbf{U}_{12}\big| }{\sqrt{1-\sin^{2}{\theta_{13}}}}=N_{2}\sin{\theta}\frac{\bigg| 1-\frac{\epsilon}{4}\left[\cos^{2}{\theta}-\left(\frac{\vert m^{0}_{1}\vert-\vert m^{0}_{3}\vert}{\vert m^{0}_{2}\vert+\vert m^{0}_{1}\vert}\right)\right]\cos^{2}{\theta}  \bigg| }{\sqrt{1-\sin^{2}{\theta_{13}}}};\nonumber\\
	\sin{\theta_{23}}&=&\frac{\big|\mathbf{U}_{23}\big|}{\sqrt{1-\sin^{2}{\theta_{13}}}}=\frac{N_{3}}{\sqrt{2}} 
	\frac{\bigg| 1-\frac{\epsilon}{4} \frac{\left[\left(\vert m^{0}_{2}\vert+\vert m^{0}_{1}\vert\right)^{2}\cos^{2}{\theta}\sin^{2}{\theta}-\left(\vert m^{0}_{2}\vert+\vert m^{0}_{1}\vert\right)\left(\vert m^{0}_{1}\vert-\vert m^{0}_{3}\vert\right)+\left(\vert m^{0}_{1}\vert-\vert m^{0}_{3}\vert\right)^{2} \right]}{\left(\vert m^{0}_{2}\vert+\vert m^{0}_{3}\vert\right)\left(\vert m^{0}_{1}\vert-\vert m^{0}_{3}\vert\right)} \bigg|}{\sqrt{1-\sin^{2}{\theta_{13}}}}.\label{cfmixa}
\end{eqnarray}

Let us consider two extreme cases where the lightest neutrino mass takes part. According to the squared mass scales $\Delta m^{2}_{21}=\vert m^{0}_{2}\vert^{2}-\vert m^{0}_{1}\vert^{2}$ and $\Delta m^{2}_{13}=\vert m^{0}_{1}\vert^{2}-\vert m^{0}_{3}\vert^{2}$, two neutrino masses might write as $\vert m^{0}_{2}\vert=\sqrt{\vert m^{0}_{1}\vert^{2}+\Delta m^{2}_{21}}$ and $\vert m^{0}_{1}\vert=\sqrt{\vert m^{0}_{3}\vert^{2}+\Delta m^{2}_{13}}$.

\begin{description}
\item [Strict inverted hierarchy ($\vert m^{0}_{3}\vert=0$)]

In this case, we have $\vert m^{0}_{1}\vert=\sqrt{\Delta m^{2}_{13}}$ and
\begin{equation}
	\vert m^{0}_{2}\vert\approx \vert m^{0}_{1}\vert\left(1+\frac{1}{2}\frac{\Delta m^{2}_{21}}{\vert m^{0}_{1}\vert^{2}}\right).
\end{equation}	
 Then, one can obtain a precise values for the mixing angles
\begin{eqnarray}
	\sin{\theta_{13}}&\approx&  \frac{ N_{3}}{4}\vert \epsilon\vert \sin{2\theta}  \bigg| \left[2(1+r_{A})\cos^{2}{\theta}-1 \right]\left(1-r_{A}\right)\bigg|=\frac{\sqrt{2}}{18} \vert \epsilon\vert(1-\frac{1}{2}r_{A});\nonumber\\
	\sin{\theta_{12}}&\approx&N_{2}\sin{\theta}\frac{\bigg| 1-\frac{\epsilon}{4}\left[\cos^{2}{\theta}-\frac{1}{2}\left(1-r_{A}\right)\right]\cos^{2}{\theta}  \bigg| }{\sqrt{1-\sin^{2}{\theta_{13}}}}=\frac{1}{\sqrt{3}}\frac{\big| 1-\frac{\epsilon}{36}\big| }{\sqrt{1-\sin^{2}{\theta_{13}}}};\nonumber\\
	\sin{\theta_{23}}&\approx&\frac{N_{3}}{\sqrt{2}} 	\frac{\bigg| 1-\frac{\epsilon}{4}\left[4\sin^{2}{\theta}\cos^{2}{\theta}-1\right]\bigg| }{\sqrt{1-\sin^{2}{\theta_{13}}}}=\frac{1}{\sqrt{2}}\frac{\big| 1+\frac{\epsilon}{36}\big| }{\sqrt{1-\sin^{2}{\theta_{13}}}},
\end{eqnarray}
where $r_{A}=\Delta m^{2}_{21}/2\Delta m^{2}_{13}$.

In the above expressions, we have considered $\sin{\theta}=1/\sqrt{3}$ which is a good approximation to the tribimaximal scenario~\cite{Harrison200376, Harrison2002167, Xing200285,Altarelli:2012ss}, $\vert \epsilon\vert=0.3$ and $\alpha_{\epsilon}=0$
we obtain $\sin{\theta_{13}}\approx0.0234$, $\sin{\theta_{12}}\approx0.587$ and $\sin{\theta_{23}}\approx0.713$. In the case where $\alpha_{\epsilon}=\pi$, the solar and atmospheric angles have similar values in comparison to above case.

\item [Almost degenerate $\vert m^{0}_{3}\vert\gg\sqrt{\Delta m^{2}_{13}}$]
In this limit, we obtain the following masses
\begin{equation}
\vert m^{0}_{1}\vert\approx \vert m^{0}_{3}\vert \left[1+r_{B}\right],\qquad \vert m^{0}_{2}\vert\approx \vert m^{0}_{3}\vert \left[1+r_{B}+r_{C}\right]
\end{equation}
with $r_{B}\approx \Delta m^{2}_{13}/2 \vert m^{0}_{3}\vert^{2}$ and $r_{C}\approx \Delta m^{2}_{21}/2 \vert m^{0}_{3}\vert^{2}$.
\end{description}

For this reason, the mixing angles formulas are written as

\begin{eqnarray}
	\sin{\theta_{13}}&\approx&  \frac{ N_{3}}{2} \sin{\theta} \cos^{3}{\theta} \frac{\vert \epsilon\vert}{r_{B}}=\frac{\sqrt{2}}{9}\frac{\vert \epsilon\vert}{r_{B}};\nonumber\\
	\sin{\theta_{12}}&\approx&N_{2}\sin{\theta}\frac{\bigg| 1-\frac{\epsilon}{4}\cos^{4}{\theta}  \bigg| }{\sqrt{1-\sin^{2}{\theta_{13}}}}=\frac{1}{\sqrt{3}}\frac{\big| 1-\frac{\epsilon}{9}\big| }{\sqrt{1-\sin^{2}{\theta_{13}}}};\nonumber\\
\sin{\theta_{23}}&\approx&\frac{N_{3}}{\sqrt{2}} 	\frac{\bigg| 1-\frac{\epsilon}{2r_{B}}\sin^{2}{\theta}\cos^{2}{\theta}\bigg| }{\sqrt{1-\sin^{2}{\theta_{13}}}}=\frac{1}{\sqrt{2}}\frac{\big| 1-\frac{\epsilon}{9 r_{B}}\big| }{\sqrt{1-\sin^{2}{\theta_{13}}}}.
\end{eqnarray}

Remarkably, in this scheme the three angles can be accommodated with great accuracy according to the experimental data as we will see later.

Before finishing this section, it is worthy of mentioning the relation among the reactor angle and the deviation of the solar and atmospheric angles, respectively. To do so, in the strict hierarchy case we have
\begin{equation}
		\vert \epsilon\vert\approx \frac{18}{\sqrt{2}}\sin{\theta_{13}},
\end{equation}
then 
\begin{equation}
	\sin{\theta_{12}}\approx\frac{1}{\sqrt{3}}\frac{\big| 1\pm\frac{\sin{\theta_{13}}}{\sqrt{8}}\big| }{\sqrt{1-\sin^{2}{\theta_{13}}}},\qquad \sin{\theta_{23}}\approx\frac{1}{\sqrt{2}}\frac{\big| 1\pm\frac{\sin{\theta_{13}}}{\sqrt{8}}\big| }{\sqrt{1-\sin^{2}{\theta_{13}}}}. 
\end{equation}

In the almost degenerate case, one can write
\begin{equation}
\vert \epsilon\vert\approx \frac{9}{\sqrt{2}}r_{B}\sin{\theta_{13}},
\end{equation}
subsequently
\begin{equation}
	\sin{\theta_{12}}\approx\frac{1}{\sqrt{3}}\frac{\big| 1\pm\frac{r_{B}}{\sqrt{2}}\sin{\theta_{13}}\big| }{\sqrt{1-\sin^{2}{\theta_{13}}}},\qquad \sin{\theta_{23}}\approx\frac{1}{\sqrt{2}}\frac{\big| 1\pm\frac{\sin{\theta_{13}}}{\sqrt{2}}\big| }{\sqrt{1-\sin^{2}{\theta_{13}}}}. 
\end{equation}
where the $\pm$ represents the $\pi$ and $0$ values for the $\alpha_{\epsilon}$ phase.

\subsubsection{Quark sector}

As we already commented, the lepton sector was studied mainly in this paper. Then, we want to address briefly the quark sector within a particular benchmark as follows. We adopted the following  alignments $\langle H_{1}\rangle=\langle H_{2}\rangle$ which is consistent with the minimization of the scalar potential~\cite{Pakvasa:1977in, Beltran:2009zz, Das:2014fea}. Hence, one gets

\begin{equation}
	{\bf \mathcal{M}}_{q}= \begin{pmatrix}
		B_{q}	& b_{q} & 	C_{q} \\
		b_{q} & A_{q} & C_{q} \\
		D_{q} 	& D_{q} & E_{q}
	\end{pmatrix},
\end{equation}
where $q=u, d$ and the defined parameters can be read of Eqn. (\ref{fmas1}). Let us remark that ${\bf \mathcal{M}}_{q}$ has bee studied exhaustively in~\cite{Canales:2013cga} and significant results were released. Nonetheless, we want to address the quark mass matrices in different way so that some assumption will be done. To do so, notice that ${\bf \mathcal{M}}_{q}$ is diagonalized \footnote{See Appendix C to more detail in the diagonalization process.} by $\mathbf{U}_{q(L,R)}$ such that $\hat{{\bf \mathcal{M}}}_{q}=\mathbf{U}^{\dagger}_{q L} \mathbf{\mathcal{M}}_{q} \mathbf{U}_{q R}$ with $\hat{{\bf \mathcal{M}}}_{q}=\textrm{Diag.}\left(m_{q_{1}}, m_{q_{2}},  m_{q_{3}}\right)$ denoting the quark physical masses. Then, the following rotations is realized $\mathbf{U}_{q(L,R)}=\mathbf{U}_{\pi/4}\mathbf{u}_{q(L, R)}$ so that  $\hat{{\bf \mathcal{M}}}_{q}=\mathbf{u}^{\dagger}_{q L} \mathbf{m}_{q} \mathbf{u}_{q R}$. Notice that

\begin{equation}
	{\bf m}_{q}= \begin{pmatrix}
		A_{q}	& b_{q} & 	0 \\
		b_{q} & B_{q} & \sqrt{2}C_{q} \\
		0 	& \sqrt{2} D_{q} & E_{q}
	\end{pmatrix},\qquad \mathbf{U}_{\pi/4}= \begin{pmatrix}
		\frac{1}{\sqrt{2}}	& \frac{1}{\sqrt{2}} & 0 \\
		-\frac{1}{\sqrt{2}} & \frac{1}{\sqrt{2}} & 0 \\
		0 	& 0 & 0
	\end{pmatrix}.
\end{equation}

At this stage, two assumptions are made $A_{q}=0$ and $C_{q}=D_{q}$. 
To be honest, we could not eliminate the former entry by means the $\mathbf{S}_{4}\otimes Z_{2}$  discrete symmetry 
and the latter assumption might be realized within the left-right theory\cite{Pati:1974yy, Mohapatra:1974gc, Senjanovic:1975rk, Senjanovic:1978ev} by invoking parity symmetry. Also, as was shown in \cite{Fritzsch:1999ee}, the second assumption can be realized by a suitable transformation in the right-handed quarks fields (there are no right-handed currents in the model), which are $\mathbf{SU(2)_{L}}$ singlets, such that the resultant quark mass matrix turns out being hermitian. Due to this fact, we could have assumed that ${\bf \mathcal{M}}_{q}$ is hermitian but only the aforementioned simplification was carried out. 
In this benchmark the quark mass matrix has the generalized Fritzsch textures~\cite{Fritzsch:1999ee, Barranco:2010we, Fritzsch:2015gxa} which fit with great accuracy the CKM mixing matrix. 

As a result, the CKM mixing matrix is given by
$\mathbf{V}=\mathbf{U}^{\dagger}_{u}\mathbf{U}_{d}=\mathbf{O}^{T}_{u}\bar{\mathbf{P}}_{q}\mathbf{O}_{d}$ where $\bar{\mathbf{P}}_{q}= \mathbf{P}^{\dagger}_{u}\mathbf{P}_{d}$ and the $\mathbf{O}_{q}$ orthogonal matrix has the following form
\begin{equation}
\mathbf{O}_{q}=\begin{pmatrix}
			\sqrt{\frac{m_{q_{3}}\vert m_{q_{2}}\vert\left(\vt E_{q}\vt-m_{q_{1}}\right)}{ R_{q_{1}}}}	& -\sqrt{\frac{m_{q_{1}} m_{q_{3}}\left(\vt E_{q}\vt+\vt m_{q_{2}}\vt\right)}{ R_{q_{2}}}} & \sqrt{\frac{m_{q_{1}}\vert m_{q_{2}}\vert\left(m_{q_{3}}-\vt E_{q}\vt\right)}{ R_{q_{3}}}} \\
			\sqrt{\frac{m_{q_{1}}\left(\vt E_{q}\vt-m_{q_{1}}\right)\vert E_{q}\vert}{ R_{q_{1}}}}	 & \sqrt{\frac{\vt m_{q_{2}}\vt \left(\vt E_{q}\vt+\vt m_{q_{2}}\vt \right)\vert E_{q}\vert}{ R_{q_{2}}}} & \sqrt{\frac{ m_{q_{3}} \left( m_{q_{3}}-\vt E_{q}\vt\right)\vert E_{q}\vert}{ R_{q_{3}}}} \\
			-\sqrt{\frac{m_{q_{1}}\left(\vt E_{q}\vt+\vt m_{q_{2}}\vt\right)\left(m_{q_{3}}-\vt E_{q}\vt\right)}{ R_{q_{1}}}}	& -\sqrt{\frac{\vt m_{q_{2}}\vt\left(\vt E_{q}\vt- m_{q_{1}}\right)\left(m_{q_{3}}-\vt E_{q}\vt\right)}{ R_{q_{2}}}} & \sqrt{\frac{ m_{q_{3}}\left(\vt E_{q}\vt- m_{q_{1}}\right)\left(\vt E_{q}\vt+\vt m_{q_{2}}\vt\right)}{ R_{q_{3}}}}
\end{pmatrix}.
\end{equation}
where $q=u,d$. In addition,
\begin{eqnarray}
	R_{q_{1}}&=&\left(m_{q_{3}}-m_{q_{1}}\right)\left(\vt m_{q_{2}}\vt+m_{q_{1}}\right)\vt E_{q}\vt;\nn\\%
	R_{q_{2}}&=&\left(m_{q_{3}}+\vt m_{q_{2}}\vt \right)\left(\vt m_{q_{2}}\vt+m_{q_{1}}\right)\vt E_{q}\vt;\nn\\
	R_{q_{3}}&=&\left(m_{q_{3}}+\vt m_{q_{2}}\vt \right)\left( m_{q_{3}}-m_{q_{1}}\right)\vt E_{q}\vt.
\end{eqnarray}

As we can show in the Appendix B,  in the CKM matrix
there are four parameters namely $\vt E_{q}\vt $ ($q=u,d$) and two effective CP-violating phases ($\alpha$ and $\beta$) so that a numerical study will be realized to fix them.

\section{Results}

\subsection{Lepton sector}
We have shown that our theoretical formulas on the mixing angles can accommodate the experimental data where the inverted hierarchy is favored. In order to get a full set of free parameter values that fit the mixing angles, then some scattered plot will be elaborated as follows.

The mixing angles depend on three free parameters, explicitly 
\begin{eqnarray}
\sin{\theta_{13}}&=&\sin{\theta_{13}}\left(\epsilon,\theta, \vert m^{0}_{3}\vert\right),\nn\\
\sin{\theta_{12}}&=&\sin{\theta_{13}}\left(\epsilon,\theta, \vert m^{0}_{3}\vert\right),\nn\\
\sin{\theta_{23}}&=&\sin{\theta_{13}}\left(\epsilon,\theta, \vert m^{0}_{3}\vert\right).
\end{eqnarray}

\begin{figure}[htp]\centering
	\includegraphics[scale=0.46]{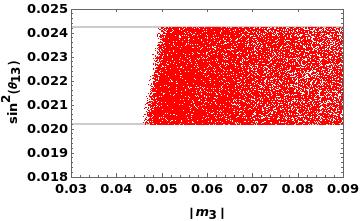}
	\hspace{1mm}\includegraphics[scale=0.46]{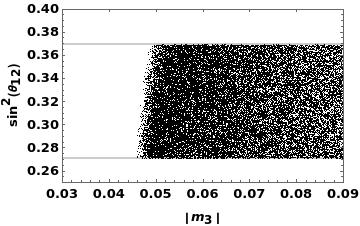}
	\includegraphics[scale=0.46]{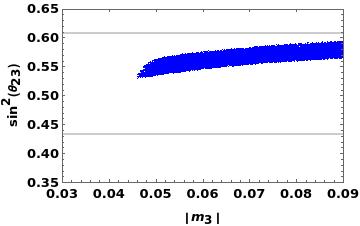}
	\caption{From left to right: the reactor, solar and atmospheric angles versus the $\vert m^{0}_{3}\vert$ lightest neutrino mass. The thick line stands for $3\sigma$ of C. L.}\label{mixanvsm3}
\end{figure}

Hence, from the previous analytical study the three free parameters let vary on the following ranges: $\epsilon \in \left[-0.3, 0\right]$, $\theta \in \left[0, \pi/3\right]$ and $\vert m^{0}_{3}\vert \in \left[0, 0.09\right]~eV$. Therefore, we
demand our theoretical formulas satisfy (at $3\sigma$)
the following values~\cite{deSalas:2020pgw}
\begin{eqnarray}
\sin^{2}{\theta_{12}}/10^{-1} &\in& 2.71-3.69, \nonumber\\ \sin^{2}{\theta_{13}}/10^{-2} &\in& 2.018-2.424,\nonumber\\
\sin^{2}{\theta_{23}}/10^{-1} &\in& 4.33-6.08,
\end{eqnarray}
for the inverted hierarchy. Additionally, 
\begin{eqnarray}
\Delta m^{2}_{21}\left[10^{-5}~eV^{2}\right] &\in& 6.94-8.14,\nonumber\\
\Delta m^{2}_{13}\left[10^{-3}~eV^{2}\right] &\in& 2.37-2.53.
\end{eqnarray}
Having included the experimental data, the scattered plots are constructed by using the theoretical formulas given in Eqn.(\ref{cfmixa}) which have to satisfy the experimental values. As a result, the mixing angles as function of the  lightest neutrino mass are displayed in Fig. (\ref{mixanvsm3}). The allowed region of values for the $\vert m^{0}_{3}\vert$ is consistent with the previous analytical study. 

As it was already commented, the $\theta$ parameter is identified with the solar angle in the limit of $\mu \leftrightarrow \tau$ exact. Then, the following scattered plots exhibit the region where $\theta$ parameter lies around the experimental value of the solar angle. In fact, this value is close to tribimaximal prediction since the solar angle receives a small correction from $\epsilon$.

\begin{figure}[htp]\centering
	\includegraphics[scale=0.46]{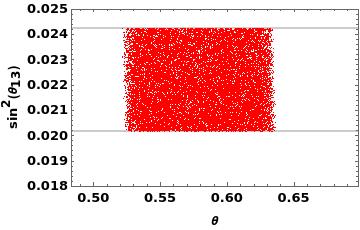}
	\hspace{1mm}\includegraphics[scale=0.46]{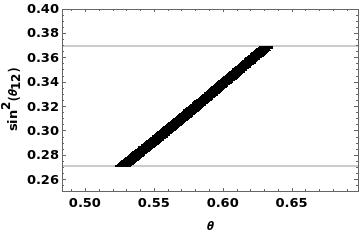}
	\includegraphics[scale=0.46]{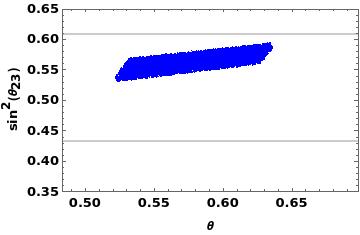}
	\caption{From left to right: the reactor, solar and atmospheric angles versus the $\theta$ parameter. The thick line stands for $3\sigma$ of C. L.}\label{mixanvsthe}
\end{figure}

\begin{figure}[htp]\centering
	\includegraphics[scale=0.46]{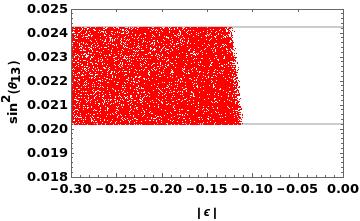}
	\hspace{1mm}\includegraphics[scale=0.46]{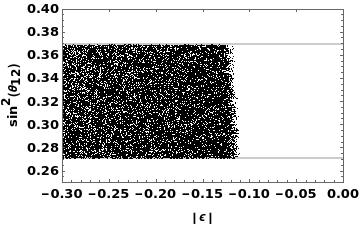}
	\includegraphics[scale=0.46]{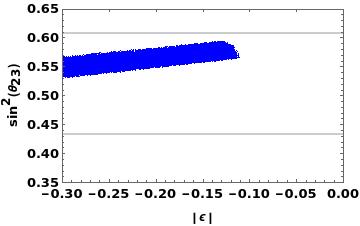}
	\caption{From left to right: the reactor, solar and atmospheric angles versus the $\vert \epsilon\vert$ parameter. The thick line stands for $3\sigma$ of C. L.}\label{mixanvseps}
\end{figure}

In the previous analytical study, we showed the $\epsilon$ parameter must be negative and this may vary in the interval $\left[0,-0.3\right]$. The numerical study shows the favored region where the mixing angles are fitted at $3\sigma$, see Fig. (\ref{mixanvseps}). Evidently, the case with $\vert \epsilon\vert=0$ is excluded due to this stands for the limit of exact $\mu \leftrightarrow\tau$ symmetry.

%\textcolor{red}{
%Additionally, we development a Least Square fitting in order to obtain the parameters best values which are given by
%\begin{eqnarray}
%\theta&=&0.582^{+0.0158}_{-0.016}\\
%\epsilon&=&-0.128^{+0.0603}_{- 0.063}\\
%m_3&=&0.0868^{+0.00207}_{-0.019}
%\end{eqnarray}
%which leads to have the following values
%\begin{eqnarray}
%\sin^2\left(\theta_{13}\right)&=&0.0225\pm 0.0024\\
%\sin^2\left(\theta_{12}\right)&=&0.3180\pm 0.0148\\
%\sin^2\left(\theta_{23}\right)&=&0.5780\pm 0.0057
%\end{eqnarray}
%}

As model prediction, we have calculated numerical the effective Majorana mass of electron neutrino  which is defined as follows
\begin{equation}
\vert m_{ee}\vert=\left| \sum^{3}_{i=1} m_{i}\mathbf{U}^{2}_{ei}	\right|
\end{equation} 

with $m_{i}$ represents the physical neutrino mass and $\mathbf{U}_{ei}$ PMNS matrix elements. This effective mass has been measured by \textbf{GERDA} phase I\cite{Agostini:2013mzu} and II~\cite{Agostini:2018tnm}, and the lowest upper bound is $\vert m_{ee}\vert<0.22~eV$.

In our model, CP parities have been used in the neutrino masses. In particular, we utilized $m^{0}_{2}=-\vert m^{0}_{2}\vert$,  $m^{0}_{1}=\vert m^{0}_{1}\vert$ and $m^{0}_{3}=\vert m^{0}_{3}\vert$ since this fit quite well the mixing angles. Consequently, the predicted region for the effective Majorana mass of electron neutrino is shown in the following scattered plots.

%\textcolor{red}{which are in accordance with our prediction on  this mass using the best fit values
%\begin{equation}
%m_{ee}=0.03769\pm 0.0030
%\end{equation}
%}

\begin{figure}[h!]\centering
	\includegraphics[scale=0.46]{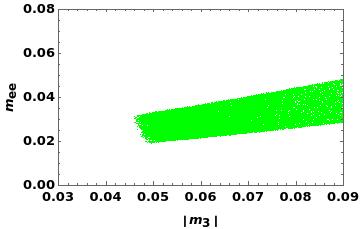}
	\hspace{1mm}\includegraphics[scale=0.46]{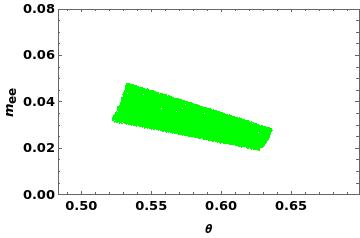}
	\includegraphics[scale=0.46]{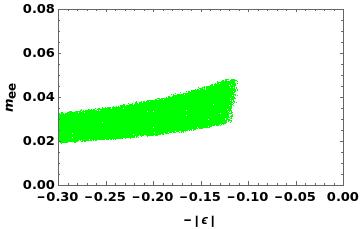}
	\caption{The effective Majorana mass of neutrino electron versus the fitted parameters.}\label{mixanvseps2}
\end{figure}

\subsection{Quark sector}
Our numerical study consists in making scattering plots. To do so, we compare our CKM theoretical expression with the standard parametrization one. In
particular, we consider the entries $\left(\mathbf{V}_{ui}\right)^{th}$ ($i=d,s, b$) and $\left(\mathbf{V}_{cb}\right)^{th}$ that depend on the free parameters
\begin{eqnarray}
	\vt\left( \mathbf{V}_{ui}\right)^{th}\vt&=&\vt\left( \mathbf{V}_{ui}\right)^{th}\left( \vt E_{u}\vt, \vt E_{d}\vt, \alpha,\beta \right)\vt;\nn\\
	\vt\left( \mathbf{V}_{cb}\right)^{th}\vt&=&\vt\left( \mathbf{V}_{ui}\right)^{th}\left( \vt E_{u}\vt, \vt E_{d}\vt, \alpha,\beta \right)\vt.
\end{eqnarray}

Therefore, we demand the magnitude of mentioned entries must satisfy the following experimental values~\cite{Workman:2022ynf}
\begin{eqnarray}
	\vt\left( \mathbf{V}_{ud}\right)^{ex}\vt&=& 0.97401\pm 0.0001;\nn\\
	\vt\left( \mathbf{V}_{us}\right)^{ex}\vt&=& 0.22650 \pm 0.0004;\nn\\
	\vt\left( \mathbf{V}_{ub}\right)^{ex}\vt&=& 0.00361^{+0.00011}_{-0.00009};\nn\\
	\vt\left( \mathbf{V}_{cb}\right)^{ex}\vt&=& 0.04053^{+0.00083}_{-0.00061}.
\end{eqnarray}

In the current study, the physical quark masses are considered as input values. To be more precise, the normalized quark masses ($m_{q_{i}}/m_{q_{3}}$) will be used due to their ratios do not change drastically at different energy scales as one can verify directly from~\cite{Xing:2007fb}.
So that, at
the top quark mass scale we have~\cite{Garces:2018nar} 
\begin{eqnarray}
	\tilde{m}_{u}&=&(1.33 \pm 0.73)\times 10^{-5},\nn\\ \tilde{m}_{c}&=&(3.91 \pm 0.42)\times 10^{-3},\nn\\ \tilde{m}_{d}&=&(1.49\pm 0.39) × 10^{-3},\nn\\ \tilde{m}_{s}&=&(2.19 \pm 0.53) \times10^{-2}.
\end{eqnarray}
In addition, for simplicity,  two dimensionless parameters have been defined $y_{q}\equiv \vt E_{q}\vt/m_{q_{3}}$ ($q=u, d$), then we now have the constraint $1>y_{q}> \tilde{m}_{q_{2}}\equiv \vt m_{q_{2}}\vt/m_{q_{3}} >\tilde{m}_{q_{1}}$. Explicitly, for the up and down sector  $1>y_{u}> \tilde{m}_{c}\equiv \vt m_{c}\vt/m_{t} >\tilde{m}_{u}$ and $1>y_{d}> \tilde{m}_{s}\equiv \vt m_{s}\vt/m_{b} >\tilde{m}_{d}$.

\begin{figure}[htp]\centering
	\includegraphics[scale=0.5]{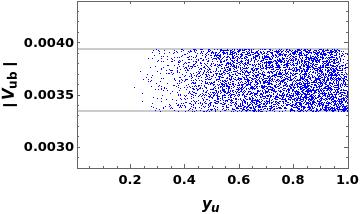}
	\hspace{1mm}\includegraphics[scale=0.5]{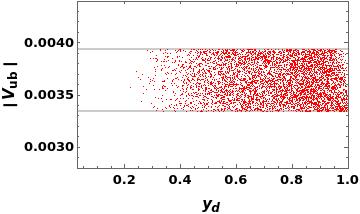}
	\includegraphics[scale=0.5]{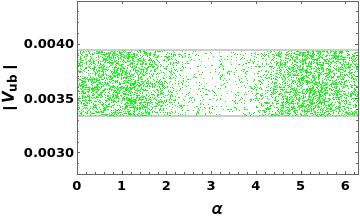}\hspace{1mm}	\includegraphics[scale=0.5]{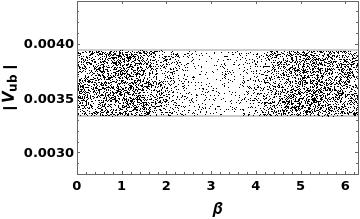}
	\caption{ $\vt V_{ub}\vt$ as function of the four free parameters. The thick line stands for $3\sigma$ of C. L.}\label{Vubvsp}
\end{figure}

With all the above information, we calculate the allowed regions for the four CKM entries and constrain the free parameter set of values. However,
let us show you only the scattered plots for $\vt V_{ub}\vt$ and $\vt V_{cb}\vt$ since that these entries usually are complicated to fit. As we already commented the theoretical expression are required to satisfy the experimental data up to $3\sigma$. Moreover, the normalized quark masses let vary up to $2\sigma$ and the two CP-violating phases are in the range $\left[0,2\pi\right]$. Then, as one notices, in Figure~\ref{Vubvsp}, there is a set of values in which $\vt V_{ub}\vt$ is fitted with great accuracy. 

\begin{figure}[htp]\centering
	\includegraphics[scale=0.5]{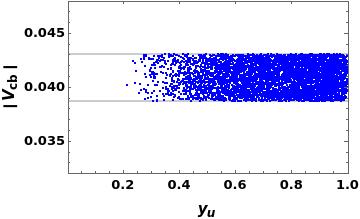}
	\hspace{1mm}\includegraphics[scale=0.5]{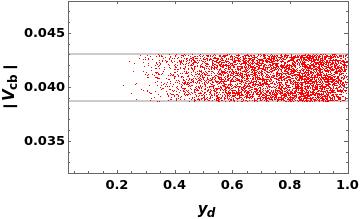}
	\includegraphics[scale=0.5]{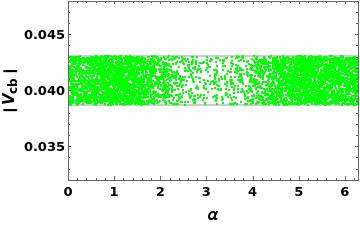}\hspace{1mm}	\includegraphics[scale=0.5]{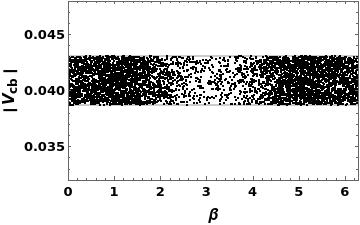}
	\caption{ $\vt V_{cb}\vt$ as function of the four free parameters. The thick line stands for $3\sigma$ of C. L.}\label{Vcbvsp}
\end{figure}

Focusing in the dimensionless parameters $y_{u}$ and $y_{d}$, the favorable regions lie in  $\left[0.5,1 \right)$ approximately. Additionally,
there are two regions of values for the CP phases, $\alpha$ and $\beta$,  where the magnitude of $\vt V_{ub}\vt$ is accommodated. In the Figure~\ref{Vcbvsp}, we see $\vt V_{cb}\vt$ as function of the four free parameters and these have the same allowed region as the above case.

To finish this section, we want to comment our naive analysis showed a large region of values for the free parameters where the theoretical CKM entries are in good agreement with the experimental data up to $3\sigma$. A strict  study, as for example an $\chi^{2}$ fit,  must determine better the space of values however the principal aim of this numerical study was shown the generalized Fritzsch mass textures fit the CKM matrix as it is well known.

\section{Summary and conclusions}
To sum up, we have built a non-renormalizable model where the fermion mixing is driven by the spontaneous breaking of the $\mathbf{S}_{4}\otimes \mathbf{Z}_{2}$ discrete group. An appropriated alignment of the scalar vev's allows to break the $\mu \leftrightarrow \tau$ symmetry in the effective neutrino mass matrix. Therefore, under a perturbative study, we were able to correct the wrong predictions on the reactor and atmospheric angles, and a set of values for the free parameters was found such that the mixing angles are consistent with the latest neutrino data. Due to the lack of extra symmetries, in the quark sector, a benchmark allows to get consistent mass textures that accommodate the CKM mixing matrix.

We have learned that the flavor symmetries have been useful to eliminate spurious parameters in the Yukawa sector. At the same time, those shape the fermion mass matrices, consequently  the mixing pattern can be obtained straightforwardly. Ambitious flavored models have gone beyond of fitting the mixings and prediction on some free parameters (Majorana phases, Dirac CP phase for instance) have been done. In conclusion, despite the $\mu \leftrightarrow \tau$ is outdated, in this constrained model, we wanted to show you that a simple soft breaking is enough to correct the mixing angles. Although the model predictions are so limited and the favored inverted hierarchy goes against the data, a soft breaking of $\mu \leftrightarrow \tau$ is still alive from theoretical point of view nevertheless the experiments have the verdict.

%\section{Conclusions}
\section*{Acknowledgements}
This work was partially supported by Secretaria de Investigación y Posgrado del Instituto Politécnico Nacional under Projects 20221285 and 20220411 and the program PAPIIT IN109321.

\appendix
\section{$\mathbf{S}_{4}$ flavour symmetry}
$\mathbf{S}_{4}$ is the group of permutations of four objects and this is the smallest non abelian group having doublet, triplet and singlet irreducible representations \cite{Ishimori:2010au}. Geometrically, $\mathbf{S}_{4}$ is the symmetry of a cube. This discrete group contains five irreducible
representations, this is, $\mathbf{1_{1},1_{2},2,3_{1},3_{2}}$ which has
the following tensor product rules \cite{Ishimori:2010au}: 
\begin{align}
	\begin{pmatrix}
		a_{1} \\ 
		a_{2}%
	\end{pmatrix}%
	_{\mathbf{2}}\otimes 
	\begin{pmatrix}
		b_{1} \\ 
		b_{2}%
	\end{pmatrix}%
	_{\mathbf{2}}& =(a_{1}b_{1}+a_{2}b_{2})_{\mathbf{1}_{1}}\oplus
	(-a_{1}b_{2}+a_{2}b_{1})_{\mathbf{1}_{2}}\oplus 
	\begin{pmatrix}
		a_{1}b_{2}+a_{2}b_{1} \\ 
		a_{1}b_{1}-a_{2}b_{2}%
	\end{pmatrix}%
	_{\mathbf{2}\ ,} \\
	\begin{pmatrix}
		a_{1} \\ 
		a_{2}%
	\end{pmatrix}%
	_{\mathbf{2}}\otimes 
	\begin{pmatrix}
		b_{1} \\ 
		b_{2} \\ 
		b_{3}%
	\end{pmatrix}%
	_{\mathbf{3}_{1}}& =%
	\begin{pmatrix}
		a_{2}b_{1} \\ 
		-\frac{1}{2}(\sqrt{3}a_{1}b_{2}+a_{2}b_{2}) \\ 
		\frac{1}{2}(\sqrt{3}a_{1}b_{3}-a_{2}b_{3})%
	\end{pmatrix}%
	_{\mathbf{3}_{1}}\oplus 
	\begin{pmatrix}
		a_{1}b_{1} \\ 
		\frac{1}{2}(\sqrt{3}a_{2}b_{2}-a_{1}b_{2}) \\ 
		-\frac{1}{2}(\sqrt{3}a_{2}b_{3}+a_{1}b_{3})%
	\end{pmatrix}%
	_{\mathbf{3}_{2}\ ,} \\
	\begin{pmatrix}
		a_{1} \\ 
		a_{2}%
	\end{pmatrix}%
	_{\mathbf{2}}\otimes 
	\begin{pmatrix}
		b_{1} \\ 
		b_{2} \\ 
		b_{3}%
	\end{pmatrix}%
	_{\mathbf{3}_{2}}& =%
	\begin{pmatrix}
		a_{1}b_{1} \\ 
		\frac{1}{2}(\sqrt{3}a_{2}b_{2}-a_{1}b_{2}) \\ 
		-\frac{1}{2}(\sqrt{3}a_{2}b_{3}+a_{1}b_{3})%
	\end{pmatrix}%
	_{\mathbf{3}_{1}}\oplus 
	\begin{pmatrix}
		a_{2}b_{1} \\ 
		-\frac{1}{2}(\sqrt{3}a_{1}b_{2}+a_{2}b_{2}) \\ 
		\frac{1}{2}(\sqrt{3}a_{1}b_{3}-a_{2}b_{3})%
	\end{pmatrix}%
	_{\mathbf{3}_{2}\ ,} \\
	\begin{pmatrix}
		a_{1} \\ 
		a_{2} \\ 
		a_{3}%
	\end{pmatrix}%
	_{\mathbf{3}_{1}}\otimes 
	\begin{pmatrix}
		b_{1} \\ 
		b_{2} \\ 
		b_{3}%
	\end{pmatrix}%
	_{\mathbf{3}_{1}}& =(a_{1}b_{1}+a_{2}b_{2}+a_{3}b_{3})_{\mathbf{1}%
		_{1}}\oplus 
	\begin{pmatrix}
		\frac{1}{\sqrt{2}}(a_{2}b_{2}-a_{3}b_{3}) \\ 
		\frac{1}{\sqrt{6}}(-2a_{1}b_{1}+a_{2}b_{2}+a_{3}b_{3})%
	\end{pmatrix}%
	_{\mathbf{2}}  \notag \\
	& \ \oplus 
	\begin{pmatrix}
		a_{2}b_{3}+a_{3}b_{2} \\ 
		a_{1}b_{3}+a_{3}b_{1} \\ 
		a_{1}b_{2}+a_{2}b_{1}%
	\end{pmatrix}%
	_{\mathbf{3}_{1}}\oplus 
	\begin{pmatrix}
		a_{3}b_{2}-a_{2}b_{3} \\ 
		a_{1}b_{3}-a_{3}b_{1} \\ 
		a_{2}b_{1}-a_{1}b_{2}%
	\end{pmatrix}%
	_{\mathbf{3}_{2}\ ,} \\
	\begin{pmatrix}
		a_{1} \\ 
		a_{2} \\ 
		a_{3}%
	\end{pmatrix}%
	_{\mathbf{3}_{2}}\otimes 
	\begin{pmatrix}
		b_{1} \\ 
		b_{2} \\ 
		b_{3}%
	\end{pmatrix}%
	_{\mathbf{3}_{2}}& =(a_{1}b_{1}+a_{2}b_{2}+a_{3}b_{3})_{\mathbf{1}%
		_{1}}\oplus 
	\begin{pmatrix}
		\frac{1}{\sqrt{2}}(a_{2}b_{2}-a_{3}b_{3}) \\ 
		\frac{1}{\sqrt{6}}(-2a_{1}b_{1}+a_{2}b_{2}+a_{3}b_{3})%
	\end{pmatrix}%
	_{\mathbf{2}}  \notag \\
	& \ \oplus 
	\begin{pmatrix}
		a_{2}b_{3}+a_{3}b_{2} \\ 
		a_{1}b_{3}+a_{3}b_{1} \\ 
		a_{1}b_{2}+a_{2}b_{1}%
	\end{pmatrix}%
	_{\mathbf{3}_{1}}\oplus 
	\begin{pmatrix}
		a_{3}b_{2}-a_{2}b_{3} \\ 
		a_{1}b_{3}-a_{3}b_{1} \\ 
		a_{2}b_{1}-a_{1}b_{2}%
	\end{pmatrix}%
	_{\mathbf{3}_{2}\ ,} \\
	\begin{pmatrix}
		a_{1} \\ 
		a_{2} \\ 
		a_{3}%
	\end{pmatrix}%
	_{\mathbf{3}_{1}}\otimes 
	\begin{pmatrix}
		b_{1} \\ 
		b_{2} \\ 
		b_{3}%
	\end{pmatrix}%
	_{\mathbf{3}_{2}}& =(a_{1}b_{1}+a_{2}b_{2}+a_{3}b_{3})_{\mathbf{1}%
		_{2}}\oplus 
	\begin{pmatrix}
		\frac{1}{\sqrt{6}}(2a_{1}b_{1}-a_{2}b_{2}-a_{3}b_{3}) \\ 
		\frac{1}{\sqrt{2}}(a_{2}b_{2}-a_{3}b_{3})%
	\end{pmatrix}%
	_{\mathbf{2}}  \notag \\
	& \ \oplus 
	\begin{pmatrix}
		a_{3}b_{2}-a_{2}b_{3} \\ 
		a_{1}b_{3}-a_{3}b_{1} \\ 
		a_{2}b_{1}-a_{1}b_{2}%
	\end{pmatrix}%
	_{\mathbf{3}_{1}}\oplus 
	\begin{pmatrix}
		a_{2}b_{3}+a_{3}b_{2} \\ 
		a_{1}b_{3}+a_{3}b_{1} \\ 
		a_{1}b_{2}+a_{2}b_{1}%
	\end{pmatrix}%
	_{\mathbf{3}_{2}\ .}
\end{align}%

In this section, we remark an interesting feature between the $\mathbf{S}_{3}$~\cite{Ishimori:2010au} and $\mathbf{S}_{4}$ non-abelian groups. As it is well known, these are different, the former one has three irreducible representations namely two singlets, $\mathbf{1_{
1}}$ and $\mathbf{1_{
2}}$, and one doublet $\mathbf{2}$. This group is smaller than $\mathbf{S}_{4}$ as one can see in ~\cite{Ishimori:2010au}. In addition, each representation of $\mathbf{S}_{4}$ can be decomposed in representation of  $\mathbf{S}_{3}$ as follows: $\mathbf{1}_{1}\rightarrow \mathbf{1}_{1}$, $\mathbf{1}_{2}\rightarrow \mathbf{1}_{2}$, $\mathbf{2}\rightarrow \mathbf{2}$, $\mathbf{3}_{1}\rightarrow \mathbf{1}_{1}\oplus \mathbf{2}$ and $\mathbf{3}_{2}\rightarrow \mathbf{1}_{2}\oplus \mathbf{2}$.

\section{Symmetry $\mu \leftrightarrow \tau$}

In the basis where the charged lepton mass matrix is diagonal, the effective neutrino mass term is given by
\begin{equation}
\mathcal{L}=\bar{\nu}_{\ell L} (\mathbf{M}_{\nu})_{\ell \ell^{\prime}}  (\nu_{\ell^{\prime}L})^{C}+h.c,
\end{equation}
where $\ell, \ell^{\prime}=e, \mu, \tau$. If the neutrino mass matrix is invariant under the interchange label  $\mu \leftrightarrow \tau$, one would have
\begin{equation}
\mathbf{M}_{\nu}=\begin{pmatrix}
m_{ee} & m_{e\mu} & m_{e\mu} \\
m_{e\mu} & m_{\mu\mu} & m_{\mu\tau} \\
m_{e\mu} & m_{\mu\tau} & m_{\mu\mu}
\end{pmatrix}.
\end{equation}

As one can notice, in the previous mass matrix the entries $12$ ($22$) and $13$ ($33$) are equals, then that matrix possesses the $\mu \leftrightarrow \tau$ symmetry \cite{Mohapatra:1998ka,Lam:2001fb,Kitabayashi:2002jd, Koide:2003rx,Fukuyama:1997ky} and its prediction on the mixing angles are obtained as follows.
In the mentioned basis, the neutrino mass matrix is diagonalized by $\mathbf{U}_{\nu}$, this means, $\mathbf{U}^{\dagger}_{\nu}\mathbf{M}_{\nu}\mathbf{U}^{\ast}_{\nu}=\hat{\mathbf{M}}$ where the latter matrix stands for the neutrino masses, $\hat{\mathbf{M}}=\textrm{Diag}.(m_{1},m_{2},m_{3})$, which can be complex.

As it is well known, $\mathbf{U}_{\nu}=\mathbf{U}_{\pi/4}\mathbf{U}_{\theta}$ so that $\mathbf{U}^{\dagger}_{\nu}\mathbf{M}_{\nu}\mathbf{U}^{\ast}_{\nu}=\mathbf{U}^{\dagger}_{\theta}\mathbf{m}_{\nu}\mathbf{U}^{\ast}_{\theta} $ where
\begin{equation}
\mathbf{m}_{\nu}=\begin{pmatrix}
	m_{ee} & \sqrt{2} m_{e\mu} & 0 \\
	\sqrt{2} m_{e\mu} & m_{\mu\mu}+m_{\mu\tau} & 0 \\
	0 & 0 & m_{\mu\mu}-m_{\mu\tau}\end{pmatrix},\qquad \mathbf{U}_{\pi/4}=\begin{pmatrix}
		1 & 0 & 0 \\
		0 & \frac{1}{\sqrt{2}} & -\frac{1}{\sqrt{2}} \\
		0 & \frac{1}{\sqrt{2}} & \frac{1}{\sqrt{2}}
	\end{pmatrix}.
\end{equation}

In addition, $\mathbf{m}_{\nu}$ is diagonalized by $\mathbf{U}_{\theta_{\nu}}$
whose form is given as
\begin{equation}
\mathbf{U}_{\theta}=\begin{pmatrix}
	\cos{\theta} & \sin{\theta} & 0 \\
	-\sin{\theta} & \cos{\theta} & 0 \\
	0 & 0 & 1
\end{pmatrix},
\end{equation}
and one condition should be satisfied
\begin{equation}
\tan{2\theta}=\frac{\sqrt{8}m_{e\mu}}{m_{\mu\mu}+m_{\mu\tau}-m_{ee}}.
\end{equation}

Therefore, one can write the full matrix
\begin{equation}
\mathbf{U}_{\nu}=\begin{pmatrix}
		\cos{\theta}  & \sin{\theta} & 0 \\
		-\frac{\sin{\theta}}{\sqrt{2}} & \frac{\cos{\theta}}{\sqrt{2}}  & -\frac{1}{\sqrt{2}} \\
		-\frac{\sin{\theta}}{\sqrt{2}} & \frac{\cos{\theta}}{\sqrt{2}}  & \frac{1}{\sqrt{2}}
	\end{pmatrix}.
\end{equation}

Comparing the above mixing matrix with the standard parametrization of the PMNS matrix, one gets the reactor and atmospheric angles are $\theta_{13}=0$ and  $\theta_{23}=\pi/4$, respectively. Speaking about the solar angle, this is free parameter and can be identified by $\theta_{12}=\theta$.

At the same time, the matrix elements can be written in terms of the physical neutrino mass as follows
\begin{eqnarray}
	m_{ee}&=&m_{1}\cos^{2}{\theta}+m_{2}\sin^{2}{\theta},\nonumber\\
	m_{e\mu}&=&\frac{\sin{2\theta}}{\sqrt{8}}\left(m_{2}-m_{1}\right),\nonumber\\
	m_{\mu\tau}&=&\frac{1}{2}\left(m_{1}\sin^{2}{\theta}+m_{2}\cos^{2}{\theta}+m_{3}\right),\nonumber\\
	m_{\mu\mu}&=&\frac{1}{2}\left(m_{1}\sin^{2}{\theta}+m_{2}\cos^{2}{\theta}-m_{3}\right).
\end{eqnarray}

\section{Perturbative study to obtain $\mathbf{U}^{\epsilon}_{\nu}$}
To start with, we have to remember the stationary perturbation theory from our lectures on quantum mechanics \cite{CohenBook}. This method is applied to a system whose Hamiltonian is given by $H=H_{0}+W$ where the eigenstates ($\phi^{0}_{n}$) and eigenvalues ($E^{0}_{n}$) of $H_{0}$ are known, also $W$ (known as a perturbation) is smaller than $H_{0}$, besides that, $H_{0}$ and $W$ are time independent. With $W=\lambda~\tilde{W}$ and $\lambda\ll 1$, at first order in the $\lambda$ perturbative parameter,  one can perform the correction to the eigenstates and eigenvalues of $H(\lambda)$ ( $H(\lambda)|\Phi_{n}(\lambda) \rangle=E_{n}(\lambda) |\Phi_{n}(\lambda) \rangle$)  which are given respectively by
\begin{equation}
|\Phi_{n}(\lambda) \rangle=|\phi^{0}_{n}\rangle +\displaystyle\sum_{k\neq n}\frac{\langle \phi^{0}_{k}| W| \phi^{0}_{n} \rangle}{E^{0}_{n}-E^{0}_{k}}|\phi^{0}_{k}\rangle, \qquad	
E_{n}(\lambda)=E^{0}_{n}+\langle \phi^{0}_{n}| W| \phi^{0}_{n} \rangle.\label{evv}
\end{equation}

Then, we adapt the above results to diagonalization problem given in the neutrino sector. As it was shown, $\mathbf{\mathcal{M}}_{\nu}$ is diagonalized by the $\mathbf{U}_{\nu}$ matrix such that
$\hat{\mathbf{M}}_{\nu}=\textrm{Diag.}\left(m_{1}, m_{2}, m_{3}\right)=\mathbf{U}^{\dagger}_{\nu}\mathbf{\mathcal{M}}_{\nu}\mathbf{U}^{\ast}_{\nu}$ with $\mathbf{U}_{\nu}=\mathbf{S}_{23}\mathbf{u}_{\nu}$, then
$\hat{\mathbf{M}}_{\nu}=\mathbf{u}^{\dagger}_{\nu}\mathbf{m}_{\nu}\mathbf{u}^{\ast}_{\nu}$. $\mathbf{m}_{\nu}$
can be written as
\begin{equation}
	\mathbf{m}_{\nu}=\overbrace{\begin{pmatrix}
			m_{ee} & m_{e\mu} & m_{e\mu} \\
			m_{e\mu} & m_{\mu \mu} & m_{\mu \tau} \\
			m_{e\mu} & m_{\mu \tau} & m_{\mu \mu}
	\end{pmatrix}}^{\mathbf{m}^{0}_{\nu}}+\overbrace{\begin{pmatrix}
			0 & 0 & 0 \\
			0 & 0 & 0 \\
			0 & 0 & m_{\mu \mu} \epsilon
	\end{pmatrix}}^{\mathbf{m}^{\epsilon}_{\nu}},
\end{equation}

where the former matrix possesses exact $\mu \leftrightarrow \tau$ symmetry and it is broken in the latter one where
the dimensionless parameter $\epsilon\equiv \left(m_{\tau \tau}-m_{\mu \mu}\right)/m_{\mu\mu}$ has been defined and this quantify the breaking. In addition, this parameter will be treated as a perturbation, thus, a pertubative study at first order in $\epsilon$ will be carried out. It was shown in the above Appendix,  $\mathbf{m}^{0}_{\nu}$ is diagonalized by $\mathbf{U}^{0}_{\nu}$, then,  $\hat{\mathbf{M}}_{\nu}=\mathbf{u}^{\dagger}_{\nu}\mathbf{m}_{\nu}\mathbf{u}^{\ast}_{\nu}$ with $\mathbf{u}_{\nu}\approx \mathbf{U}^{0}_{\nu} \mathbf{U}^{\epsilon}_{\nu}$ implies
\begin{equation}
	\hat{\mathbf{M}}_{\nu}=\mathbf{U}^{\epsilon \dagger}_{\nu}\left[ \mathbf{U}^{0\dagger}_{\nu} \mathbf{m}^{0}_{\nu}\mathbf{U}^{0\ast}_{\nu} +\mathbf{U}^{0\dagger}_{\nu} \mathbf{m}^{\epsilon}_{\nu}\mathbf{U}^{0\ast}_{\nu}\right]\mathbf{U}^{\epsilon \ast}_{\nu},\quad \textrm{with} \quad \mathbf{U}^{0\dagger}_{\nu} \mathbf{m}^{0}_{\nu}\mathbf{U}^{0\ast}_{\nu}=\textrm{Diag.}\left(m^{0}_{1}, m^{0}_{2}, m^{0}_{3}\right).
\end{equation}

In addition,
\begin{equation}
\overbrace{\mathbf{U}^{0\dagger}_{\nu} \mathbf{m}^{\epsilon}_{\nu}\mathbf{U}^{0\ast}_{\nu}}^{\tilde{m}^{\epsilon}_{\nu}}=
	\frac{\epsilon~m^{0}_{\mu\mu}}{2}\begin{pmatrix}
		\sin^{2}{\theta} &-\frac{\sin{2\theta}}{2} & -\sin{\theta}\\
		-\frac{\sin{2\theta}}{2} & \cos^{2}{\theta} & \cos{\theta} \\
		-\sin{\theta}	& \cos{\theta} & 1
	\end{pmatrix},\qquad m^{0}_{\mu\mu}=\frac{1}{2}\left(m^{0}_{1}\sin^{2}{\theta}+m^{0}_{2}\cos^{2}{\theta}-m^{0}_{3}\right).
\end{equation}
In here, we make contact with the perturbation theory, instead of having a Hamiltonian, we have $\mathbf{m}_{\nu}=\mathbf{m}^{0}_{\nu}+\mathbf{m}^{\epsilon}_{\nu}$ where the eigenstates ($\nu^{0}_{i}$) and eigenvalues ($m^{0}_{i}$) of $\mathbf{m}^{0}_{\nu}$ are well known. In consequence, 
$\mathbf{U}^{\epsilon}_{\nu}$ is built by means the adapted eigenvector, this is,  $\mathbf{U}^{\epsilon}_{\nu}=\left(|\Phi_{1}(\epsilon) \rangle, |\Phi_{2}(\epsilon) \rangle, |\Phi_{3}(\epsilon) \rangle\right)$ where 

\begin{equation}
|\Phi_{n}(\epsilon) \rangle=|\nu^{0}_{n}\rangle +\displaystyle\sum^{3}_{k\neq n}\frac{\langle \nu^{0}_{k}| \tilde{m}^{\epsilon}_{\nu}| \nu^{0}_{n} \rangle}{m^{0}_{n}-m^{0}_{k}}|\nu^{0}_{k}\rangle
\end{equation}
where $n=1,2,3$. In order to apply correctly the formula, we have to observe that matrix $\mathbf{m}^{\epsilon}_{\nu}$ has been rotated by $\mathbf{U}^{0}_{\nu}$ so that the new perturbative matrix is denoted by $\tilde{m}^{\epsilon}_{\nu}$.
Additionally, each eigenvector $|\Phi_{n}(\epsilon) \rangle$ must be normalized. Finally, one gets
\begin{equation}
	\mathbf{U}^{\epsilon}_{\nu}\approx \begin{pmatrix}
		N_{1} & 	-\frac{m^{0}_{\mu\mu}}{m^{0}_{2}-m^{0}_{1}}\frac{\sin{2\theta}}{4}\epsilon N_{2} & \frac{m^{0}_{\mu\mu}}{m^{0}_{1}-m^{0}_{3}}\frac{\sin{\theta}}{2}\epsilon N_{3}\\
		\frac{m^{0}_{\mu\mu}}{m^{0}_{2}-m^{0}_{1}}\frac{\sin{2\theta}}{4}\epsilon N_{1} & N_{2}  & -\frac{m^{0}_{\mu\mu}}{m^{0}_{2}-m^{0}_{3}}\frac{\cos{\theta}}{2}\epsilon N_{3}  \\
		-\frac{m^{0}_{\mu\mu}}{m^{0}_{1}-m^{0}_{3}}\frac{\sin{\theta}}{2}\epsilon N_{1} & \frac{m^{0}_{\mu\mu}}{m^{0}_{2}-m^{0}_{3}}\frac{\cos{\theta}}{2}\epsilon N_{2}  & N_{3}
	\end{pmatrix},
\end{equation}
where the normalization factors are written as
\begin{eqnarray}
	N_{1}&=&\left[1+\frac{\vert \epsilon\vert^{2} \sin^{2}{\theta}}{4}\left(\bigg|\frac{m^{0}_{\mu\mu}}{m^{0}_{1}-m^{0}_{3}} \bigg|^{2}+\cos^{2}{\theta}\bigg| \frac{m^{0}_{\mu\mu}}{m^{0}_{2}-m^{0}_{1}} \bigg|^{2}\right)\right]^{-1/2},\nonumber\\
	N_{2}&=&\left[1+\frac{\vert \epsilon\vert^{2} \cos^{2}{\theta}}{4}\left(\bigg|\frac{m^{0}_{\mu\mu}}{m^{0}_{2}-m^{0}_{3}} \bigg|^{2}+\sin^{2}{\theta}\bigg|\frac{m^{0}_{\mu\mu}}{m^{0}_{2}-m^{0}_{1}} \bigg|^{2}\right)\right]^{-1/2},\nonumber\\
	N_{3}&=&\left[1+\frac{\vert \epsilon\vert^{2} }{4}\left(\sin^{2}{\theta}\bigg|\frac{m^{0}_{\mu\mu}}{m^{0}_{1}-m^{0}_{3}} \bigg|^{2}+\cos^{2}{\theta}\bigg|\frac{m^{0}_{\mu\mu}}{m^{0}_{2}-m^{0}_{3}} \bigg|^{2}\right)\right]^{-1/2}.
\end{eqnarray}

To sum up, $\mathbf{\mathcal{M}}_{\nu}$ is diagonalized approximately by $\mathbf{U}_{\nu}\approx\mathbf{S}_{23}\mathbf{u}_{\nu}$ with $\mathbf{u}_{\nu}\approx \mathbf{U}^{0}_{\nu}\mathbf{U}^{\epsilon}_{\nu}$. As a result of this, the PMNS mixing matrix, $\mathbf{U}$, is defined as $\mathbf{U}=\mathbf{U}^{\dagger}_{e L} \mathbf{U}_{\nu}$ where $\mathbf{U}_{e L}=\mathbf{S}_{23}\mathbf{u}_{e L}$ was performed in the lepton section. Therefore,  $\mathbf{U}\approx \mathbf{u}^{\dagger}_{e L} \mathbf{U}^{0}_{\nu}\mathbf{U}^{\epsilon}_{\nu}$, as one realizes, $\mathbf{u}^{\dagger}_{e L}$ contains unphysical phases which are irrelevant to the magnitude for PMNS matrix elements that are written explicitly as
\begin{eqnarray}
\left( \mathbf{U}\right)_{11}&=&N_{1}\cos{\theta}\left[1+\frac{\epsilon}{2}\sin^{2}{\theta}\left(\frac{m^{0}_{\mu\mu}}{m^{0}_{2}-m^{0}_{1}}\right)\right];\nn\\
\left(\mathbf{U}\right)_{12}&=&N_{2}\sin{\theta}\left[1-\frac{\epsilon}{2}\cos^{2}{\theta}\left(\frac{m^{0}_{\mu\mu}}{m^{0}_{2}-m^{0}_{1}}\right)\right];\nn\\
\left(\mathbf{U}\right)_{13}&=&\frac{N_{3}}{4}\sin{2\theta}~ \epsilon~\left[\frac{m^{0}_{\mu\mu}\left(m^{0}_{2}-m^{0}_{1}\right)}{\left(m^{0}_{2}-m^{0}_{3}\right)\left(m^{0}_{1}-m^{0}_{3}\right)}\right];\nn\\
\left(\mathbf{U}\right)_{21}&=&-\frac{N_{1}}{\sqrt{2}}\sin{\theta}\left[1-\frac{\epsilon}{2} m^{0}_{\mu\mu}\left\{\frac{1}{m^{0}_{1}-m^{0}_{3}}+\frac{\cos^{2}{\theta}}{m^{0}_{2}-m^{0}_{1}}\right\}\right];\nn\\
\left(\mathbf{U}\right)_{22}&=&\frac{N_{2}}{\sqrt{2}}\cos{\theta}\left[1-\frac{\epsilon}{2} m^{0}_{\mu\mu}\left\{\frac{1}{m^{0}_{2}-m^{0}_{3}}-\frac{\sin^{2}{\theta}}{m^{0}_{2}-m^{0}_{1}}\right\}\right];\nn\\
\left(\mathbf{U}\right)_{23}&=&-\frac{N_{3}}{\sqrt{2}}\left[1-\frac{\epsilon}{2} m^{0}_{\mu\mu}\left\{\frac{\cos^{2}{\theta}}{m^{0}_{2}-m^{0}_{3}}+\frac{\sin^{2}{\theta}}{m^{0}_{1}-m^{0}_{3}}\right\}\right];\nn\\
\left(\mathbf{U}\right)_{31}&=&-\frac{N_{1}}{\sqrt{2}}\sin{\theta}\left[1+\frac{\epsilon}{2} m^{0}_{\mu\mu}\left\{\frac{1}{m^{0}_{1}-m^{0}_{3}}-\frac{\cos^{2}{\theta}}{m^{0}_{2}-m^{0}_{1}}\right\}\right];\nn\\
\left(\mathbf{U}\right)_{32}&=&\frac{N_{2}}{\sqrt{2}}\cos{\theta}\left[1+\frac{\epsilon}{2} m^{0}_{\mu\mu}\left\{\frac{1}{m^{0}_{2}-m^{0}_{3}}+\frac{\sin^{2}{\theta}}{m^{0}_{2}-m^{0}_{1}}\right\}\right];\nn\\
\left(\mathbf{U}\right)_{33}&=&\frac{N_{3}}{\sqrt{2}}\left[1-\frac{\epsilon}{2} m^{0}_{\mu\mu}\left\{\frac{\cos^{2}{\theta}}{m^{0}_{2}-m^{0}_{3}}+\frac{\sin^{2}{\theta}}{m^{0}_{1}-m^{0}_{3}}\right\}\right].
\end{eqnarray}

In the limit $\epsilon\rightarrow 0$, one recovers the $\mu \leftrightarrow \tau$ scenario.

\section{Diagonalizing $\mathbf{\mathcal{M}}_{q}$}
After the spontaneous symmetry breaking and with the appropriated alignment in the vev's, the quark mass term is given by 
\begin{equation}
	\bar{q}_{L}\mathbf{\mathcal{M}}_{q}q_{R}+h.c.
\end{equation}
with $q=u,d$. Explicit, we have
\begin{equation}
	\mathbf{\mathcal{M}}_{q}=\begin{pmatrix}
		B_{q}	& b_{q} & 	C_{q} \\
		b_{q} & A_{q} & C_{q} \\
		D_{q} 	& D_{q} & E_{q}
	\end{pmatrix}.
\end{equation}
In order to diagonalize the above mass matrix \footnote{To see more details in the diagonalization method, we suggest to read \cite{CanalesTeDoc}}, a transformation in the left and right-handed quarks is made such that $q_{L}=\mathbf{U}_{q L}~\tilde{q}_{L}$ and $q_{R}=\mathbf{U}_{q R}~\tilde{q}_{R}$ where $\tilde{q}_{(L, R)}$ are the quark fields in the mass basis. Then, $\hat{{\bf \mathcal{M}}}_{q}=\mathbf{U}^{\dagger}_{q L} {\bf \mathcal{M}}_{q} \mathbf{U}_{q R}$ where $\hat{{\bf \mathcal{M}}}_{q}=\textrm{Diag.}(m_{q_{1}}, m_{q_{2}}, m_{q_{3}})$ stands for the quark physical masses. With $\mathbf{U}_{q L}=\mathbf{U}_{\pi/4}\mathbf{u}_{q(L, R)}$, one obtains $\hat{{\bf \mathcal{M}}}_{q}=\mathbf{u}^{\dagger}_{q L} {\bf m}_{q} \mathbf{u}_{q R}$, in this case, we have
\begin{equation}
	{\bf m}_{q}= \begin{pmatrix}
		A_{q}	& b_{q} & 	0 \\
		b_{q} & B_{q} & \sqrt{2}C_{q} \\
		0 	& \sqrt{2} D_{q} & E_{q}
	\end{pmatrix},\qquad \mathbf{U}_{\pi/4}= \begin{pmatrix}
		\frac{1}{\sqrt{2}}	& \frac{1}{\sqrt{2}} & 0 \\
		-\frac{1}{\sqrt{2}} & \frac{1}{\sqrt{2}} & 0 \\
		0 	& 0 & 0
	\end{pmatrix}.
\end{equation}

As it was discussed, a benchmark was considered such that $A_{q}=0$ and $C_{q}=D_{q}$ so that we end up having a complex symmetric mass matrix

\begin{equation}
	{\bf m}_{q}= \begin{pmatrix}
		0	& b_{q} & 	0 \\
		b_{q} & B_{q} & \sqrt{2}C_{q} \\
		0 	& \sqrt{2} C_{q} & E_{q}
	\end{pmatrix}.
\end{equation}
Given ${\bf m}_{q}$, this can be written in the polar form, this means, $b_{q}=\vert b_{q}\vert e^{i \alpha_{b_{q}}}$, $B_{q}=\vert B_{q}\vert e^{i \alpha_{B_{q}}}$ and so forth. The phases can be absorbed in the quark fields, to do so let us write ${\bf m}_{q}=\mathbf{P}_{q} \bar{{\bf m}} \mathbf{P}_{q}$ with $\mathbf{P}_{q}=\textrm{Diag.}\left(e^{i\eta_{q_{1}}}, e^{i\eta_{q_{2}}}, e^{i\eta_{q_{3}}} \right)$ where the following condition must be satisfied
\begin{equation}
	\eta_{q_{1}}=\frac{2\textrm{arg}(b_{q})-\textrm{arg}(B_{q})}{2},\qquad \eta_{q_{2}}=\frac{\textrm{arg}(B_{q})}{2},\qquad \eta_{q_{3}}=\frac{\textrm{arg}(E_{q})}{2},\qquad \textrm{arg}(B_{q})+\textrm{arg}(E_{q})=2 \textrm{arg}(\sqrt{2}~C_{q});
\end{equation}
and 
\begin{equation}
	\bar{{\bf m}}_{q}= \begin{pmatrix}
		0	&\vt  b_{q}\vt & 	0 \\
		\vt	b_{q} \vt & \vt B_{q}\vt & \vt\sqrt{2}C_{q} \vt\\
		0 	& \vt \sqrt{2} C_{q}\vt & \vt E_{q}\vt
	\end{pmatrix}.\label{qmmwf}
\end{equation}

As a result of this, $\mathbf{u}_{q L}=\mathbf{P}_{q}\mathbf{O}_{q}$ and $\mathbf{u}_{q R}=\mathbf{P}^{\dagger}_{q}\mathbf{O}_{q}$, $\mathbf{O}_{q}$ being the orthogonal matrix that diagonalizes to $\bar{{\bf m}}_{q}$. Therefore, $\hat{{\bf \mathcal{M}}}_{q}=\mathbf{u}^{\dagger}_{q L} {\bf m}_{q} \mathbf{u}_{q R}= \mathbf{O}^{T}_{q} \bar{{\bf m}}_{q} \mathbf{O}_{q}$, this last expression is useful to fix three free parameters, in terms of the quark physical masses and one unfixed parameter ($\vt E_{q}\vt$), through the following invariants
\begin{equation}
	\textrm{tr} \left( \hat{{\bf \mathcal{M}}}_{q} \right), \qquad \textrm{tr} \left( \hat{{\bf \mathcal{M}}}^{2}_{q} \right),\qquad \textrm{det} \left( \hat{{\bf \mathcal{M}}}_{q} \right)
\end{equation}
where tr and det stand for the trace and determinant. In consequence, 
\begin{equation}
	\vt  b_{q}\vt=\sqrt{\frac{m_{q_{3}}\vert m_{q_{2}}\vert m_{q_{1}}}{\vt E_{q}\vt}},\qquad \vt B_{q}\vt=  m_{q_{3}}-\vt m_{q_{2}}\vt+m_{q_{1}}-\vt E_{q}\vt,\qquad \sqrt{2}\vt C_{q}\vt=\sqrt{\frac{m_{q_{3}}\vert m_{q_{2}}\vert m_{q_{1}}}{\vt E_{q}\vt}}.
\end{equation}
In the above parameters, $m_{q_{2}}=-\vert m_{q_{2}}\vert$ has been chosen in order to have real parameters.

Once many free parameters have been fixed, the $\mathbf{O}_{q}$ orthogonal matrix 
is built by means the $X_{q_{i}}$ eigenvectors, $\mathbf{O}_{q}=\left(X_{q_{1}},-X_{q_{2}}, X_{q_{3}} \right)$,  which are given by
\begin{equation}
	X_{q_{i}}=\frac{1}{N_{q_{i}}}\begin{pmatrix}
		\vt  b_{q}\vt \vt\sqrt{2}C_{q} \vt	\\
		m_{q_{i}} \vt\sqrt{2}C_{q} \vt	\\
		m_{q_{i}}\left(m_{q_{i}}-\vt  B_{q}\vt \right)	-\vt  b_{q}\vt^{2} 
	\end{pmatrix}
\end{equation}
Here, $N_{q_{i}}$ stands for the normalization factors whose explicit form is determined by the condition $X^{T}_{q_{i}}X_{q_{i}}=1$. 
Finally, one obtains
\begin{equation}
	\mathbf{O}_{q}=\begin{pmatrix}
		\sqrt{\frac{m_{q_{3}}\vert m_{q_{2}}\vert\left(\vt E_{q}\vt-m_{q_{1}}\right)}{ R_{q_{1}}}}	& -\sqrt{\frac{m_{q_{1}} m_{q_{3}}\left(\vt E_{q}\vt+\vt m_{q_{2}}\vt\right)}{ R_{q_{2}}}} & \sqrt{\frac{m_{q_{1}}\vert m_{q_{2}}\vert\left(m_{q_{3}}-\vt E_{q}\vt\right)}{ R_{q_{3}}}} \\
		\sqrt{\frac{m_{q_{1}}\left(\vt E_{q}\vt-m_{q_{1}}\right)\vert E_{q}\vert}{ R_{q_{1}}}}	 & \sqrt{\frac{\vt m_{q_{2}}\vt \left(\vt E_{q}\vt+\vt m_{q_{2}}\vt \right)\vert E_{q}\vert}{ R_{q_{2}}}} & \sqrt{\frac{ m_{q_{3}} \left( m_{q_{3}}-\vt E_{q}\vt\right)\vert E_{q}\vert}{ R_{q_{3}}}} \\
		-\sqrt{\frac{m_{q_{1}}\left(\vt E_{q}\vt+\vt m_{q_{2}}\vt\right)\left(m_{q_{3}}-\vt E_{q}\vt\right)}{ R_{q_{1}}}}	& -\sqrt{\frac{\vt m_{q_{2}}\vt\left(\vt E_{q}\vt- m_{q_{1}}\right)\left(m_{q_{3}}-\vt E_{q}\vt\right)}{ R_{q_{2}}}} & \sqrt{\frac{ m_{q_{3}}\left(\vt E_{q}\vt- m_{q_{1}}\right)\left(\vt E_{q}\vt+\vt m_{q_{2}}\vt\right)}{ R_{q_{3}}}}
	\end{pmatrix}.
\end{equation}
In addition
\begin{eqnarray}
	R_{q_{1}}&=&\left(m_{q_{3}}-m_{q_{1}}\right)\left(\vt m_{q_{2}}\vt+m_{q_{1}}\right)\vt E_{q}\vt;\nn\\
	R_{q_{2}}&=&\left(m_{q_{3}}+\vt m_{q_{2}}\vt \right)\left(\vt m_{q_{2}}\vt+m_{q_{1}}\right)\vt E_{q}\vt;\nn\\
	R_{q_{3}}&=&\left(m_{q_{3}}+\vt m_{q_{2}}\vt \right)\left( m_{q_{3}}-m_{q_{1}}\right)\vt E_{q}\vt.
\end{eqnarray}

Let us point out the unfixed parameter, $\vt E_{q}\vt$, has to satisfy the constraint $m_{q_{3}}>\vt E_{q}\vt>\vt m_{q_{2}}\vt>m_{q_{1}}$ in order to get a real orthogonal matrix, $\mathbf{O}_{q}$.

Having calculated the ingredients that take places in the CKM matrix, we have $\mathbf{U}_{q L}=\mathbf{U}_{\pi/4}\mathbf{u}_{q L}=\mathbf{U}_{\pi/4}\mathbf{P}_{q}\mathbf{O}_{q}$.  Consequently, 
$\mathbf{V}=\mathbf{U}^{\dagger}_{u L}\mathbf{U}_{d L}=\mathbf{O}^{T}_{u}\bar{\mathbf{P}}_{q}\mathbf{O}_{d}$ with $\bar{\mathbf{P}}_{q}= \mathbf{P}^{\dagger}_{u}\mathbf{P}_{d}$, Here, let us emphasize an important point, this has to do with the CP phases that enter in the CKM matrix. Notice that $\bar{\mathbf{P}}_{q}$ contains three phases but two of them only play an important role to fit the mixings. At the end, $\mathbf{V}$
has four parameters namely $\vt E_{q}\vt $ ($q=u,d$) and two effective CP-violating phases ($\alpha$ and $\beta$).

In the CKM matrix, the involved matrices  are written explicitly
\begin{eqnarray}
	\mathbf{O}_{u}&=&\begin{pmatrix}
		\sqrt{\frac{m_{t}\vert m_{c}\vert\left(\vt E_{u}\vt-m_{u}\right)}{ R_{u}}}	& -\sqrt{\frac{m_{u} m_{t}\left(\vt E_{u}\vt+\vt m_{c}\vt\right)}{ R_{c}}} & \sqrt{\frac{m_{u}\vert m_{c}\vert\left(m_{t}-\vt E_{u}\vt\right)}{ R_{t}}} \\
		\sqrt{\frac{m_{u}\left(\vt E_{u}\vt-m_{u}\right)\vert E_{u}\vert}{ R_{u}}}	 & \sqrt{\frac{\vt m_{c}\vt \left(\vt E_{u}\vt+\vt m_{c}\vt \right)\vert E_{u}\vert}{ R_{c}}} & \sqrt{\frac{ m_{t} \left( m_{t}-\vt E_{u}\vt\right)\vert E_{u}\vert}{ R_{t}}} \\
		-\sqrt{\frac{m_{u}\left(\vt E_{u}\vt+\vt m_{c}\vt\right)\left(m_{t}-\vt E_{u}\vt\right)}{ R_{u}}}	& -\sqrt{\frac{\vt m_{c}\vt\left(\vt E_{u}\vt- m_{u}\right)\left(m_{t}-\vt E_{u}\vt\right)}{ R_{c}}} & \sqrt{\frac{ m_{t}\left(\vt E_{u}\vt- m_{u}\right)\left(\vt E_{u}\vt+\vt m_{c}\vt\right)}{ R_{t}}}
	\end{pmatrix};\nn\\
	\mathbf{O}_{d}&=&\begin{pmatrix}
		\sqrt{\frac{m_{b}\vert m_{s}\vert\left(\vt E_{d}\vt-m_{d}\right)}{ R_{d}}}	& -\sqrt{\frac{m_{d} m_{b}\left(\vt E_{d}\vt+\vt m_{s}\vt\right)}{ R_{s}}} & \sqrt{\frac{m_{d}\vert m_{s}\vert\left(m_{b}-\vt E_{d}\vt\right)}{ R_{b}}} \\
		\sqrt{\frac{m_{d}\left(\vt E_{d}\vt-m_{d}\right)\vert E_{d}\vert}{ R_{d}}}	 & \sqrt{\frac{\vt m_{s}\vt \left(\vt E_{d}\vt+\vt m_{s}\vt \right)\vert E_{d}\vert}{ R_{s}}} & \sqrt{\frac{ m_{b} \left( m_{b}-\vt E_{d}\vt\right)\vert E_{d}\vert}{ R_{b}}} \\
		-\sqrt{\frac{m_{d}\left(\vt E_{d}\vt+\vt m_{s}\vt\right)\left(m_{b}-\vt E_{d}\vt\right)}{ R_{d}}}	& -\sqrt{\frac{\vt m_{s}\vt\left(\vt E_{d}\vt- m_{d}\right)\left(m_{b}-\vt E_{d}\vt\right)}{ R_{s}}} & \sqrt{\frac{ m_{b}\left(\vt E_{d}\vt- m_{d}\right)\left(\vt E_{d}\vt+\vt m_{s}\vt\right)}{ R_{b}}}
	\end{pmatrix};\nn\\
	\bar{\mathbf{P}}_{q}&=&\textrm{Diag}\left(e^{i\bar{\eta}_{q_{1}}}, e^{i\bar{\eta}_{q_{2}}}, e^{i\bar{\eta}_{q_{3}}}\right).
\end{eqnarray}
with $\bar{\eta}_{q_{1}}=\eta_{d}-\eta_{u}$, $\bar{\eta}_{q_{2}}=\eta_{s}-\eta_{c}$ and  $\bar{\eta}_{q_{3}}=\eta_{t}-\eta_{b}$. In addition,
\begin{eqnarray}
	R_{u}=\left(m_{t}-m_{u}\right)\left(\vt m_{c}\vt+m_{u}\right)\vt E_{u}\vt,\,
	R_{c}=\left(m_{t}+\vt m_{c}\vt \right)\left(\vt m_{c}\vt+m_{u}\right)\vt E_{u}\vt,\,
	R_{t}=\left(m_{t}+\vt m_{c}\vt \right)\left( m_{t}-m_{u}\right)\vt E_{u}\vt;\nn\\
	R_{d}=\left(m_{b}-m_{d}\right)\left(\vt m_{s}\vt+m_{d}\right)\vt E_{d}\vt,\,
	R_{s}=\left(m_{b}+\vt m_{s}\vt \right)\left(\vt m_{s}\vt+m_{d}\right)\vt E_{d}\vt,\,
	R_{b}=\left(m_{b}+\vt m_{s}\vt \right)\left( m_{b}-m_{d}\right)\vt E_{d}\vt.	
\end{eqnarray}

On the other hand, for the up and down quark sector we have the following constraints on the free parameters $m_{t}>\vt E_{u}\vt>\vt m_{c}\vt>m_{u}$ and $m_{b}>\vt E_{d}\vt>\vt m_{s}\vt>m_{d}$. Having written the main ingredients that take place in the quark mixing, the CKM matrix elements are
\begin{eqnarray}
	\mathbf{V}^{ud}_{CKM}= \left(\mathbf{O}_{u}\right)_{11}\left(\mathbf{O}_{d}\right)_{11}e^{i\bar{\eta}_{q_{1}}}+\left(\mathbf{O}_{u}\right)_{21}\left(\mathbf{O}_{d}\right)_{21}e^{i\bar{\eta}_{q_{2}}}+\left(\mathbf{O}_{u}\right)_{31}\left(\mathbf{O}_{d}\right)_{31}e^{i\bar{\eta}_{q_{3}}};\nn\\
	\mathbf{V}^{us}_{CKM}= \left(\mathbf{O}_{u}\right)_{11}\left(\mathbf{O}_{d}\right)_{12}e^{i\bar{\eta}_{q_{1}}}+\left(\mathbf{O}_{u}\right)_{21}\left(\mathbf{O}_{d}\right)_{22}e^{i\bar{\eta}_{q_{2}}}+\left(\mathbf{O}_{u}\right)_{31}\left(\mathbf{O}_{d}\right)_{32}e^{i\bar{\eta}_{q_{3}}};\nn\\
	\mathbf{V}^{ub}_{CKM}= \left(\mathbf{O}_{u}\right)_{11}\left(\mathbf{O}_{d}\right)_{13}e^{i\bar{\eta}_{q_{1}}}+\left(\mathbf{O}_{u}\right)_{21}\left(\mathbf{O}_{d}\right)_{23}e^{i\bar{\eta}_{q_{2}}}+\left(\mathbf{O}_{u}\right)_{31}\left(\mathbf{O}_{d}\right)_{33}e^{i\bar{\eta}_{q_{3}}};\nn\\
	\mathbf{V}^{cd}_{CKM}= \left(\mathbf{O}_{u}\right)_{12}\left(\mathbf{O}_{d}\right)_{11}e^{i\bar{\eta}_{q_{1}}}+\left(\mathbf{O}_{u}\right)_{22}\left(\mathbf{O}_{d}\right)_{21}e^{i\bar{\eta}_{q_{2}}}+\left(\mathbf{O}_{u}\right)_{32}\left(\mathbf{O}_{d}\right)_{31}e^{i\bar{\eta}_{q_{3}}};\nn\\
	\mathbf{V}^{cs}_{CKM}= \left(\mathbf{O}_{u}\right)_{12}\left(\mathbf{O}_{d}\right)_{12}e^{i\bar{\eta}_{q_{1}}}+\left(\mathbf{O}_{u}\right)_{22}\left(\mathbf{O}_{d}\right)_{22}e^{i\bar{\eta}_{q_{2}}}+\left(\mathbf{O}_{u}\right)_{32}\left(\mathbf{O}_{d}\right)_{32}e^{i\bar{\eta}_{q_{3}}};\nn\\
	\mathbf{V}^{cb}_{CKM}= \left(\mathbf{O}_{u}\right)_{12}\left(\mathbf{O}_{d}\right)_{13}e^{i\bar{\eta}_{q_{1}}}+\left(\mathbf{O}_{u}\right)_{22}\left(\mathbf{O}_{d}\right)_{23}e^{i\bar{\eta}_{q_{2}}}+\left(\mathbf{O}_{u}\right)_{32}\left(\mathbf{O}_{d}\right)_{33}e^{i\bar{\eta}_{q_{3}}};\nn\\
	\mathbf{V}^{td}_{CKM}= \left(\mathbf{O}_{u}\right)_{13}\left(\mathbf{O}_{d}\right)_{11}e^{i\bar{\eta}_{q_{1}}}+\left(\mathbf{O}_{u}\right)_{23}\left(\mathbf{O}_{d}\right)_{21}e^{i\bar{\eta}_{q_{2}}}+\left(\mathbf{O}_{u}\right)_{33}\left(\mathbf{O}_{d}\right)_{31}e^{i\bar{\eta}_{q_{3}}};\nn\\
	\mathbf{V}^{ts}_{CKM}= \left(\mathbf{O}_{u}\right)_{13}\left(\mathbf{O}_{d}\right)_{12}e^{i\bar{\eta}_{q_{1}}}+\left(\mathbf{O}_{u}\right)_{23}\left(\mathbf{O}_{d}\right)_{22}e^{i\bar{\eta}_{q_{2}}}+\left(\mathbf{O}_{u}\right)_{33}\left(\mathbf{O}_{d}\right)_{32}e^{i\bar{\eta}_{q_{3}}};\nn\\
	\mathbf{V}^{tb}_{CKM}= \left(\mathbf{O}_{u}\right)_{13}\left(\mathbf{O}_{d}\right)_{13}e^{i\bar{\eta}_{q_{1}}}+\left(\mathbf{O}_{u}\right)_{23}\left(\mathbf{O}_{d}\right)_{23}e^{i\bar{\eta}_{q_{2}}}+\left(\mathbf{O}_{u}\right)_{33}\left(\mathbf{O}_{d}\right)_{33}e^{i\bar{\eta}_{q_{3}}}.
\end{eqnarray} 
Remarkable, for each entry its magnitude only depends on two effective phases, this is, $\alpha_{q}\equiv \bar{\eta}_{q_{2}}- \bar{\eta}_{q_{1}}$ and  $\beta_{q}\equiv \bar{\eta}_{q_{3}}- \bar{\eta}_{q_{1}}$.

%\bibliographystyle{utphys}
%\bibliography{BiblioQ4}

\bibliographystyle{bib_style_T1}
\bibliography{references.bib}

%\begin{thebibliography}{9}
%	\input{biblioS4.tex}
%\end{thebibliography}

\end{document}